\documentclass[useAMS,usenatbib]{mn2e}

\usepackage{color}
\usepackage{colortbl}
\usepackage{multirow}
\usepackage{dcolumn}
\usepackage{amsmath}
\usepackage{amssymb}
\usepackage{graphicx}
\usepackage{epsfig}
\usepackage{changebar}
\usepackage{natbib}
\usepackage{multirow}
\usepackage{subfigure}
\usepackage{txfonts}
\usepackage{rotating}

\usepackage{longtable,lscape}

\newcolumntype{d}[1]{D{.}{\cdot}{#1}}

\newcolumntype{.}{D{.}{.}{-1}}

\newcommand{\msun}{M$_\odot$}

\newcommand{\mclump}{\emph{M}$_{\rm{clump}}$}

\newcommand{\vlsr}{V$_{\rm{LSR}}$}

\newcommand{\mum}{$\umu$m}

\newcommand{\kms}{km\,s$^{-1}$}

\newcommand{\hi}{H~{\sc i}}

\newcommand{\hii}{H~{\sc ii}}

\newcommand{\poi}{Poisson}

\newcommand{\sex}{\texttt{SExtractor}}

\newcommand{\rms}{r.m.s.}
\newcommand{\submm}{submillimetre}

\newcommand{\KS}{Kolmogorov-Smirnov}

\newcommand{\Scu}{Scutum-Centaurus}
\newcommand{\Sag}{Sagittarius}
\newcommand{\Per}{Perseus}

\title[Dusty environments of methanol masers]{The \emph{almost} ubiquitous association of 6.7\,GHz methanol masers with dust\thanks{The full version of Figs. 3 and 5 are only available in electronic form of the journal while the full versions of Tables 1, 2 and 4 will only be available through CDS via anonymous ftp to cdsarc.u-strasbg.fr (130.79.125.5) or via http://cdsweb.u-strasbg.fr/cgi-bin/qcat?J/MNRAS/.}}

\author[J. S. Urquhart et al.]{J.\,S.\,Urquhart$^{1}$\thanks{E-mail:
jurquhart@mpifr-bonn.mpg.de (MPIfR)}, T.\,J.\,T.\,Moore$^{2}$, K.\,M.\,Menten$^{1}$, C.\,K\"onig$^{1}$, F.\,Wyrowski$^{1}$, M.\,A.\,Thompson$^{3}$, \newauthor   T.\,Csengeri$^{1}$, S.\,Leurini$^{1}$, D.\,J.\,Eden$^{4,2}$  \\
$^{1}$
 Max-Planck-Institut f\"ur Radioastronomie, Auf dem H\"ugel
  69, Bonn, Germany\\
   $^{2}$Astrophysics Research Institute, Liverpool John Moores University, IC2, Liverpool Science Park, 146 Brownlow Hill, Liverpool, L3\,5RF, UK \\
  $^{3}$Centre for Astrophysics Research, Science and Technology Research Institute, University of Hertfordshire, College Lane, Hatfield, AL10 9AB, UK \\
  $^{4}$Observatoire astronomique de Strasbourg, Universit\'e de Strasbourg, CNRS, UMR 7550, 11 rue de l'Universit\'e, 67000, Strasbourg, France
  }

\begin{document}

\date{Accepted ??. Received ??; in original form ??}

\pagerange{\pageref{firstpage}--\pageref{lastpage}} \pubyear{2009}

\maketitle

\label{firstpage}

\begin{abstract}

We report the results of 870-\mum continuum observations, using the Large APEX Bolometer Camera (LABOCA), towards 77 class-{\bf II}, 6.7-GHz methanol masers identified by the Methanol Multibeam (MMB) survey to map the thermal emission from cool dust towards these objects. These data complement a study of 630 methanol masers associated with compact dense clumps identified from the ATLASGAL survey. Compact dust emission is detected towards 70 sources, which implies a dust-association rate of 99\,per\,cent for the full MMB catalogue. Evaluation of the derived dust and maser properties leads us to conclude that the combined sample represents a single population tracing the same phenomenon. We find median clump masses of a few 10$^3$\,\msun\ and that all but a handful of sources satisfy the mass-size criterion required for massive star formation. This study provides the strongest evidence of the almost ubiquitous association of methanol masers with massive, star-forming clumps. The fraction of methanol-maser associated clumps is a factor of $\sim$2 lower in the outer Galaxy than the inner Galaxy, possibly a result of the lower metallicity environment of the former. We find no difference in the clump-mass and maser-luminosity distributions of the inner and outer Galaxy. The maser-pumping and clump-formation mechanisms are therefore likely to be relatively invariant to Galactic location. Finally, we use the ratio of maser luminosity and clump mass to investigate the hypothesis that the maser luminosity is a good indicator of the evolutionary stage of the embedded source, however, we find no evidence to support this.

\end{abstract}

\section{Introduction}
\label{sect:intro}

Feedback from massive stars and the processes associated with their formation and evolution (e.g., radiation, molecular outflows, stellar winds and supernova explosions)  play a crucial role in the structure and evolution of their host galaxy \citep{kennicutt2005}. Since star formation is exclusively associated with molecular clouds, it is important to understand how different environmental conditions can affect their properties and structure, and how these in turn can affect star formation.

The first detections of the 6.7-GHz, class-{\bf II} methanol masers (hereafter referred to as ``methanol masers" or just ``masers") were reported by \citet{menten1991}. He searched for this transition towards 123 molecular clouds, most of which showed signs of massive star formation (e.g., H$_2$O and/or OH masers and luminous IRAS sources) and detected emission from 80 regions. \citet{walsh2003} searched for dust emission towards 71 methanol masers at 450 and 850\,\mum\ using SCUBA on the JCMT. This study detected emission towards all but one source (i.e., G10.10+0.73), and this was later found to be a spurious maser detection (\citealt{breen2013}), resulting in a 100\,per\,cent detection rate for \citet{walsh2003}. A study of methanol masers located in the southern Galactic plane was conducted by \citet{hill2005} with similarly high detections rates and many of the following studies have assumed a ubiquitous link between massive star formation and the presence of methanol masers. However, these have often been based on small samples (e.g., \citealt{minier2003,breen2013}) and therefore the ubiquitous association of methanol masers with dust emission and their exclusive association with massive star formation has still not been robustly demonstrated.

Methanol masers have been found to be \emph{mostly} associated with the earliest stages of massive star formation (e.g., \citealt{minier2003,urquhart2014b}) and therefore provide a convenient and straightforward way of identifying a large sample of young embedded \emph{massive} stars in a range of different environments across the Galaxy. The Methanol Multibeam (MMB) survey (\citealt{green2009_full}) has identified 703 methanol masers between $186\degr < \ell < 20\degr$ and $|b| < 2\degr$ with a further 4 masers being found towards the Orion-Monoceros Complex (\citealt{green2012}). The full published catalogue therefore consists of 707 methanol masers (\citealt{caswell2010b,green2010,caswell2011,green2012}). Distances have been determined for the majority of these masers from the radial velocity of the strongest intensity maser component and the \citet{reid2009} Galactic rotation model.  Archival \hi\ data has been used to solve the near/far kinematic-distance ambiguity inherent to all sources located within the Solar circle (\citealt{green2011b}). 

The ATLASGAL survey has mapped 420 square degrees of the inner Galaxy (\citealt{schuller2009}; $280\degr < \ell < 60\degr$ and $|b| < 1.5\degr$) at a wavelength of 870\,\mum\ where it is sensitive to the thermal emission from cold dust. This survey has identified $\sim$10,000 dense clumps located throughout the inner Galactic plane. The ATLASGAL Compact Source Catalogue (CSC; \citealt{contreras2013,urquhart2014c}) is $\sim$98\,per\,cent complete above 5$\sigma$ ($\sim$300\,mJy\,beam$^{-1}$) and is sensitive to all compact ($<2\arcmin$) massive clumps ($>$1000\,\msun) to a heliocentric distance of 20\,kpc. 

There are 671 methanol masers located in the region common to both the ATLASGAL and MMB surveys (i.e., $280\degr < \ell < 20\degr$ and $|b| < 1.5\degr$). In a previous study we correlated the positions of these masers with the ATLASGAL CSC (\citealt{urquhart2013a}; hereafter Paper\,I) and were able to positionally associate 630 methanol masers with 577 dense clumps. Combining distances calculated by \citet{green2011b} with additional analysis presented in Paper\,I we were able to resolve the heliocentric distances to $\sim$500 of these ATLASGAL-MMB associated clumps and estimate their dust properties, masses and Galactic distribution. 

Curiously, this study failed to identify the host clumps for 41 methanol masers located within the ATLASGAL region. However, inspection of the mid-infrared images and dust-emission maps revealed that many of these masers do appear to be associated with either mid-infrared extinction, dust lanes or diffuse 8-\mum\ emission, all of which are often found towards star-forming regions. Furthermore, there is evidence of weak or diffuse low surface brightness dust emission towards a large number of these undetected sources. It is therefore likely that they are indeed star-forming regions but that their associated dust emission falls below the threshold required for detection and inclusion in the ATLASGAL CSC. However, for a small number of methanol masers, there is no evidence of association with star formation in the mid-infrared images or any dust emission, weak or otherwise. We speculated that these masers may be associated with more evolved stars and, although this was considered rather unlikely, we were unable to discount this possibility without additional data.

In order to investigate the nature of these 41 methanol masers undetected in ATLASGAL, we have conducted more sensitive submillimetre-continuum observations with the APEX telescope.\footnote{This publication is based on data acquired with the Atacama Pathfinder Experiment (APEX). APEX is a collaboration between the Max-Planck-Institut fur Radioastronomie, the European Southern Observatory, and the Onsala Space Observatory.} In addition, we observed 36 masers located outside the region of the Galactic plane surveyed by ATLASGAL (i.e., $280\degr < \ell < 20\degr$ and $|b| > 1.5\degr$, or $180\degr < \ell < 280\degr$). Although the number of methanol masers located outside the ATLASGAL region is relatively small ($\sim$5\,per\,cent of the MMB catalogue) their inclusion in the sample is important as they constitute approximately 50\,per\,cent of the methanol masers found outside the Solar circle.

Combined with the ATLASGAL-MMB associations already established, these additional observations provide a complete census of the dust properties and environments of massive star forming clumps associated with methanol masers. This large representative sample covers a significant fraction of the Galactic plane and Galactocentric distances (2-14\,kpc) and therefore provides robust statistical results that are applicable to the whole Galactic population. We use this sample to make statistically reliable comparisons of physical properties of the methanol masers and their host clumps as a function of Galactic location.

The structure of this paper is as follows: in Sect.\,2 we describe the observations and data reduction procedures. We also briefly describe the source extraction and consistency tests used to check the reliability of the catalogue produced. In Sect.\,3 we present the results of the  observations, while in Sect.\,4 we discuss the methods used to determine distances to the new associations. Also in Sect.\,4 we estimate the physical properties of the methanol masers and their host clumps and briefly discuss their distributions. In Sect.\,5 we discuss the properties of the new associations with respect to the previous analysis and investigate the nature of the methanol masers not associated with dust emission. We also discuss the Galactic distribution of the whole sample and compare the properties of the inner and outer Galaxy populations, and test the hypothesis that the methanol maser luminosity can be used as a reliable indicator of the evolutionary state of the embedded protostellar object. We present a summary of our main findings in Sect.\,6. 

\begin{figure*}
\begin{center}
\includegraphics[width=0.98\textwidth, trim= 0 0 0 0]{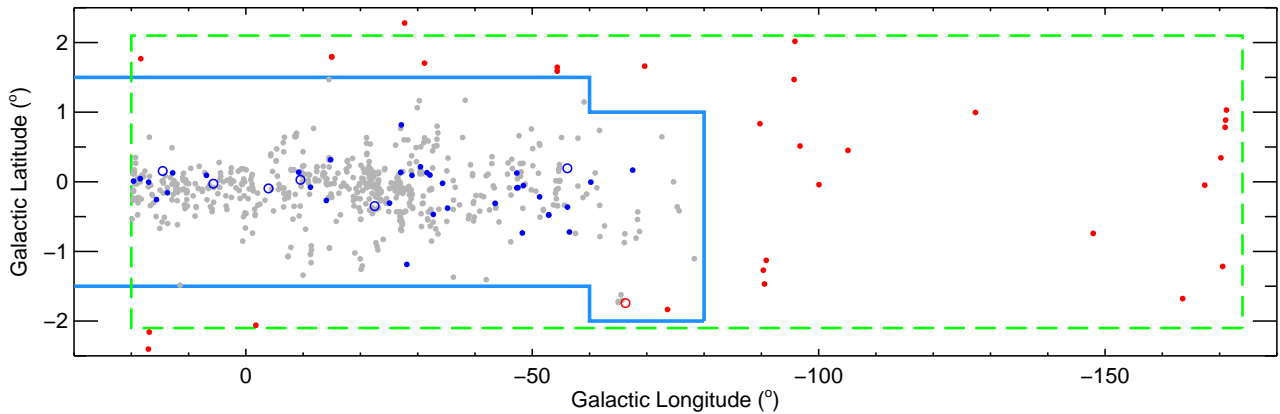}

\caption{\label{fig:lb_distribution} Galactic distribution of all MMB sources located between $186\degr < \ell < 20\degr$ and $|b|<2\degr$. The regions enclosed by the green and blue lines show the areas covered by the MMB and ATLASGAL surveys, respectively. Methanol masers matched with an ATLASGAL source are shown as filled grey circles. The blue circles indicate positions within the ATLASGAL survey region where the longer integration times were used and the red circles indicate locations outside the ATLASGAL region where shorter integration times were used. The open circles show the positions of methanol masers towards which no \submm\ emission has been detected. The two sources within the ATLASGAL area shown as red circles were mistakenly not included in the analysis presented in Paper\,I and have been observed as part of this project. Note that the Orion-Monoceros complex is not shown as it is located $\sim$20\degr\ away from the mid-plane.} 

\end{center}
\end{figure*}

\section{APEX observations and data reduction}

To complement the sample of inner Galaxy methanol maser associated clumps identified in Paper\,I, we have made \submm\ continuum observations towards 77 MMB sources that either lie outside the region covered by the ATLASGAL survey (i.e., $180\degr < \ell < 280\degr$ or $|b| > 1.5\degr$) or were not found to be associated with an ATLASGAL source. Fig.\,\ref{fig:lb_distribution} shows the Galactic distribution of the MMB sources and indicates those observed as part of this project.

These observations were made with LABOCA (\citealt{siringo2009}) between 26 and 31 August, and 31 October and 4 November 2012. The observing conditions were good with precipitable water vapour (PWV) measurements between 0.7 and 2.6\,mm with a median value of $\sim$1\,mm. A region of 6-10\arcmin\ diameter was mapped, centred on the position of the methanol maser. Sky dips were made at the beginning of each observing session and repeated typically every few hours to measure the atmospheric optical depth.  Observations of a planet and secondary calibrators were also made on a similar time scale for the purposes of absolute flux calibration and to correct the telescope pointing. The data reduction package BoA (\citealt{schuller2012}) was used to process the observations using standard procedures.

\begin{figure}
\begin{center}
\includegraphics[width=0.49\textwidth, trim= 0 0 0 0]{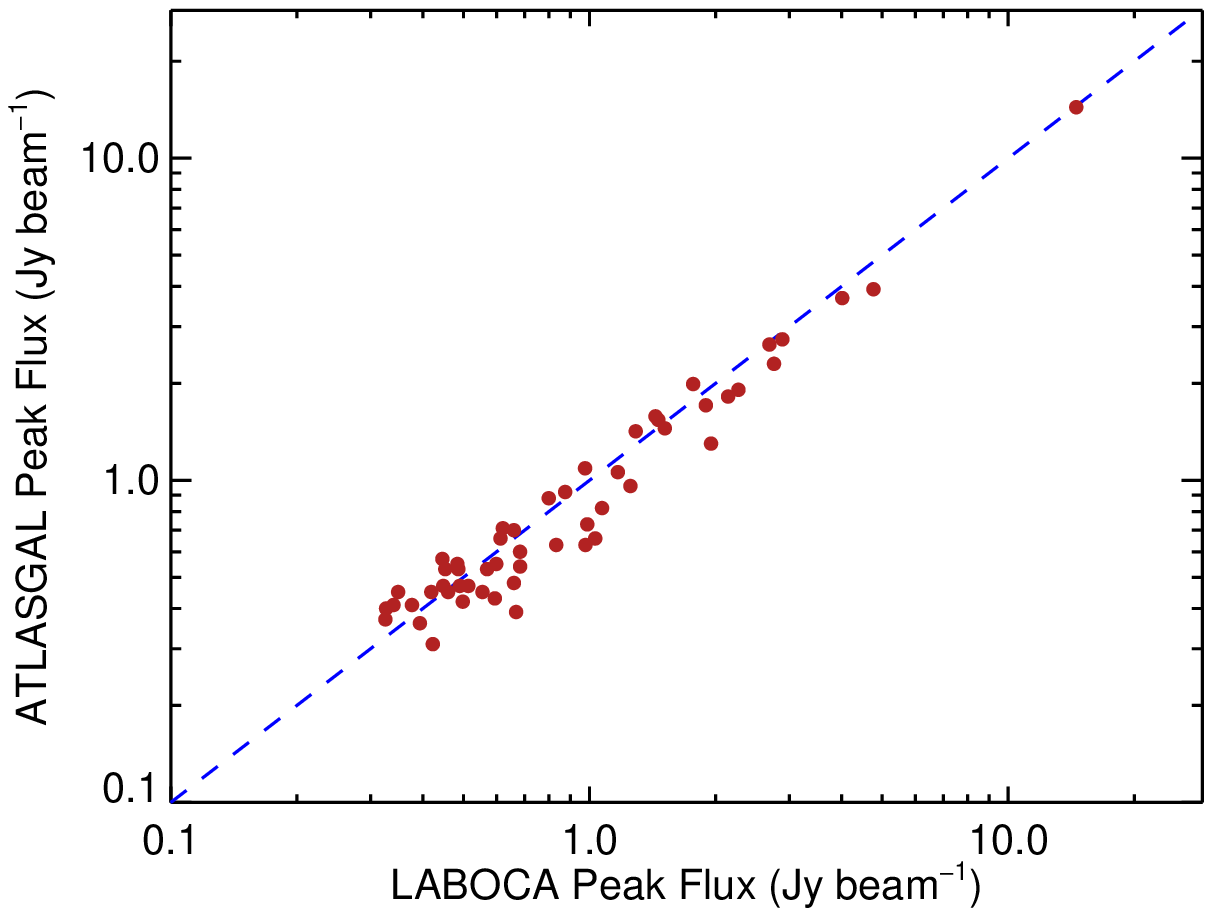}

\caption{\label{fig:flux_comparison} Comparison of the peak flux densities for 53 sources that are common to the ATLASGAL CSC and the observations reported here. The dashed blue line shows the line of equality. } 

\end{center}
\end{figure}

\setlength{\tabcolsep}{6pt}

\begin{table*}


\begin{center}\caption{Observed field positions and map sensitivity.}
\label{tbl:obs_parameters}
\begin{minipage}{\linewidth}
\small
\begin{tabular}{lcccccc}
\hline \hline
  \multicolumn{1}{c}{Field name}&  \multicolumn{1}{c}{RA}&	\multicolumn{1}{c}{Dec.}  &\multicolumn{1}{c}{Integration time}&\multicolumn{1}{c}{PWV}&\multicolumn{1}{c}{\rms}  &\multicolumn{1}{c}{ATLASGAL} \\
    \multicolumn{1}{c}{ }&  \multicolumn{1}{c}{(J2000)}&	\multicolumn{1}{c}{(J2000)}  &\multicolumn{1}{c}{(\arcmin)} &\multicolumn{1}{c}{(mm)}&\multicolumn{1}{c}{(mJy\,beam$^{-1}$)} &\multicolumn{1}{c}{flag$^{\rm{a}}$}\\
\hline
G005.665$-$00.056	&	17:58:44.96	&	$-$24:05:25.5	&	16.8	&	0.9	&	39	&	yes	\\
G006.869+00.058	&	18:00:55.87	&	$-$22:59:22.7	&	17.0	&	0.9	&	45	&	yes	\\
G012.776+00.169	&	18:12:48.61	&	$-$17:46:39.2	&	16.8	&	0.7	&	54	&	yes	\\
G013.686$-$00.118	&	18:15:41.27	&	$-$17:06:55.0	&	16.9	&	0.6	&	63	&	yes	\\
G014.499+00.128	&	18:16:24.06	&	$-$16:17:00.9	&	16.9	&	0.9	&	39	&	yes	\\
G015.586$-$00.245	&	18:19:54.74	&	$-$15:30:09.7	&	16.9	&	0.7	&	42	&	yes	\\
G016.865$-$02.108	&	18:29:13.45	&	$-$15:14:34.7	&	17.0	&	0.9	&	59	&	no	\\
G016.973+00.041	&	18:21:34.33	&	$-$14:08:42.2	&	39.2	&	2.6	&	64	&	yes	\\
G017.024$-$02.354	&	18:30:25.89	&	$-$15:12:56.3	&	5.7	&	2.8	&	172	&	no	\\
G018.348+01.815	&	18:17:48.70	&	$-$12:05:42.6	&	5.6	&	2.8	&	127	&	no	\\
\hline\\
\end{tabular}\\
$^{\rm{a}}$ This column indicates whether a field was inside or outside the ATLASGAL survey region.\\ 
Notes: Only a small portion of the data is provided here, the full table is available in electronic form at the CDS via anonymous ftp to cdsarc.u-strasbg.fr (130.79.125.5) or via http://cdsweb.u-strasbg.fr/cgi-bin/qcat?J/MNRAS/.

\end{minipage}

\end{center}
\end{table*}

\setlength{\tabcolsep}{6pt}

In total, 72 fields were observed with five fields containing two masers (i.e., G189.030+00.783, G208.996$-$19.386 --- Orion~A, G305.634+01.645, G307.132$-$00.476 and G345.012+01.797); 32 fields towards the methanol masers located  outside the ATLASGAL region with typical integration times of $\sim$6 minutes, and 40 fields located within the ATLASGAL region where longer integration times of $\sim$15-30 minutes were used in order to improve on the sensitivity of ATLASGAL. The mean map sensitivities achieved for the shorter and longer integration times is $\sim$142$\pm$78 and 51$\pm$8\,mJy \rms, respectively, and a standard deviation of $\sim$10\,mJy\,beam$^{-1}$. We note that the mean sensitivity obtained by the deeper integrations is only marginally better than that of ATLASGAL (i.e., survey average $\sim$60\,mJy\,beam$^{-1}$). However, many of these sources are located in complex regions where the ATLASGAL sensitivity is poorer and therefore the observations reported here represent a factor of two sensitivity improvement. To improve the signal-to-noise ratio (SNR) in the maps and place a tighter constraint on the upper limit we have co-added both the ATLASGAL and our maps together, however, the improvement is rather modest ($\sim$20\,per\,cent) and so this has only been done for the non-detections. We present the observational parameters for the observed fields in Table\,\ref{tbl:obs_parameters}.

The source extraction was preformed using the \sex\ algorithm using the same method and input parameters used to produce the ATLASGAL CSC (see \citealt{contreras2013} for details). In total 303 sources were detected with an average of 3 \submm\ sources found in each field. Of these, we find 53 that are located within the ATLASGAL region that were previously detected and reported in the ATLASGAL CSC. We use these to check our calibration by comparing the measured fluxes with those previously reported (Fig.\,\ref{fig:flux_comparison}). We find good agreement, with an average peak flux density ratio of 1.09$\pm$0.03 and a standard deviation of $\sim$20\,per\,cent; the absolute flux uncertainty of the ATLASGAL survey is $\sim$15\,per\,cent (\citealt{schuller2009}).

In Table\,\ref{tbl:cattable}, we present the extracted source parameters for the 303 \submm\ sources identified. This table has the same format and columns as the ATLASGAL CSC to facilitate comparisons between the two samples. We note that, in 13 of these sources, \sex\ was unable to determine a reliable integrated flux due to the complexity of the emission. In Fig.\,\ref{fig:irac_images_uchiis} we present examples of the reduced emission maps for three of the observed fields (the full version of this figure will be available in the online version of the paper).

\setlength{\tabcolsep}{6pt}

\begin{table*}


\begin{center}\caption{\label{tbl:cattable} The \submm\ sources detected in the reduced LABOCA maps.  The columns are as follows: (1) name derived from Galactic coordinates of the maximum intensity in the source; (2)-(3) Galactic coordinates of maximum intensity in the catalogue source; (4)-(5) Galactic coordinates of emission centroid; (6)-(8) semi-major and semi-minor size and source position angle measured anti-clockwise from Galactic north; (9) effective radius of source; (10)-(13) peak and integrated flux densities and their associated uncertainties; (14) \sex\ detection flag (see \citep{bertin1996} notes on these flags); (15) signal to noise ratio (SNR) --- values for sources with peak flux below $6\sigma$ detection should not be used blindly.}
\begin{minipage}{\linewidth}
\begin{tabular}{l....rrrr....c.}
  \hline \hline
  \multicolumn{1}{c}{Catalogue}
  &  \multicolumn{1}{c}{$\ell_{\mathrm{max}}$} &  \multicolumn{1}{c}{$b_{\mathrm{max}}$}
  &  \multicolumn{1}{c}{$\ell$} &  \multicolumn{1}{c}{$b$} &
  \multicolumn{1}{c}{$\sigma_{\rm{maj}}$} &  \multicolumn{1}{c}{$\sigma_{\rm{min}}$} &  \multicolumn{1}{c}{PA} &
  \multicolumn{1}{c}{$\theta_{R_{\rm{eff}}}$} &  \multicolumn{1}{c}{$S_{\rm{peak}}$} & \multicolumn{1}{c}{$\Delta S_{\rm{peak}}$} & \multicolumn{1}{c}{$S_{\rm{int}}$}& \multicolumn{1}{c}{$\Delta S_{\rm{int}}$} & \multicolumn{1}{c}{Flag} & \multicolumn{1}{c}{SNR} \\
  
  \multicolumn{1}{c}{Name} &  \multicolumn{1}{c}{($^{\circ}$)} &
  \multicolumn{1}{c}{($^{\circ}$)} &  \multicolumn{1}{c}{($^{\circ}$)} &
  \multicolumn{1}{c}{($^{\circ}$)} &  \multicolumn{1}{c}{($''$)}
  &  \multicolumn{1}{c}{($''$)} &  \multicolumn{1}{c}{($^{\circ}$)}&  \multicolumn{1}{c}{($''$)}
  &  \multicolumn{2}{c}{(Jy\,beam$^{-1}$)} &  \multicolumn{2}{c}{(Jy)}&\\
 
  \multicolumn{1}{c}{(1)} &  \multicolumn{1}{c}{(2)} &  \multicolumn{1}{c}{(3)} &  \multicolumn{1}{c}{(4)} &
  \multicolumn{1}{c}{(5)} &  \multicolumn{1}{c}{(6)} &  \multicolumn{1}{c}{(7)} &  \multicolumn{1}{c}{(8)} &
  \multicolumn{1}{c}{(9)} &  \multicolumn{1}{c}{(10)} &  \multicolumn{1}{c}{(11)} & \multicolumn{1}{c}{(12)} &  \multicolumn{1}{c}{(13)} &  \multicolumn{1}{c}{(14)} &  \multicolumn{1}{c}{(15)} \\
  \hline
G005.618$-$00.081	&	5.618	&	-0.081	&	5.616	&	-0.082	&	38	&	20	&	173	&	62.5	&	2.76	&	0.04	&	19.58	&	1.46	&	2	&	70.1	\\
G005.660$-$00.024	&	5.660	&	-0.024	&	5.659	&	-0.024	&	14	&	12	&	162	&	23.9	&	0.35	&	0.04	&	1.85	&	0.51	&	0	&	8.9	\\
G006.923+00.039	&	6.923	&	0.039	&	6.924	&	0.040	&	15	&	7	&	123	&	$\cdots$	&	0.39	&	0.05	&	1.59	&	0.44	&	0	&	8.7	\\
G006.881+00.093	&	6.881	&	0.093	&	6.882	&	0.092	&	14	&	9	&	128	&	18.0	&	0.47	&	0.05	&	1.98	&	0.53	&	0	&	10.4	\\
G013.678$-$00.050	&	13.678	&	-0.050	&	13.671	&	-0.051	&	22	&	10	&	95	&	27.7	&	0.98	&	0.06	&	6.51	&	0.91	&	0	&	15.6	\\
G013.713$-$00.084	&	13.713	&	-0.084	&	13.715	&	-0.085	&	29	&	22	&	107	&	57.7	&	1.95	&	0.06	&	17.54	&	1.43	&	2	&	31.2	\\
G013.735$-$00.082	&	13.735	&	-0.082	&	13.734	&	-0.082	&	20	&	10	&	88	&	26.2	&	1.07	&	0.06	&	5.44	&	0.77	&	3	&	17.1	\\
G013.753$-$00.123	&	13.753	&	-0.123	&	13.754	&	-0.122	&	21	&	12	&	129	&	30.7	&	0.43	&	0.06	&	4.95	&	0.86	&	2	&	7.0	\\
G013.701$-$00.126	&	13.701	&	-0.126	&	13.699	&	-0.127	&	24	&	17	&	17	&	43.9	&	0.39	&	0.06	&	4.04	&	0.81	&	3	&	6.2	\\
G013.691$-$00.158	&	13.691	&	-0.158	&	13.693	&	-0.154	&	18	&	15	&	12	&	33.1	&	0.35	&	0.06	&	1.63	&	0.53	&	2	&	5.6	\\
  \hline
\end{tabular}\\
\end{minipage}
Notes: Only a small portion of the data is provided here, the full table is only  available in electronic form at the CDS via anonymous ftp to cdsarc.u-strasbg.fr (130.79.125.5) or via http://cdsweb.u-strasbg.fr/cgi-bin/qcat?J/MNRAS/.
\end{center}
\end{table*}

\setlength{\tabcolsep}{6pt}

\subsection{Detection and matching statistics}

\begin{figure*}
\begin{center}

\includegraphics[width=0.49\textwidth, trim= 0 0 0 0]{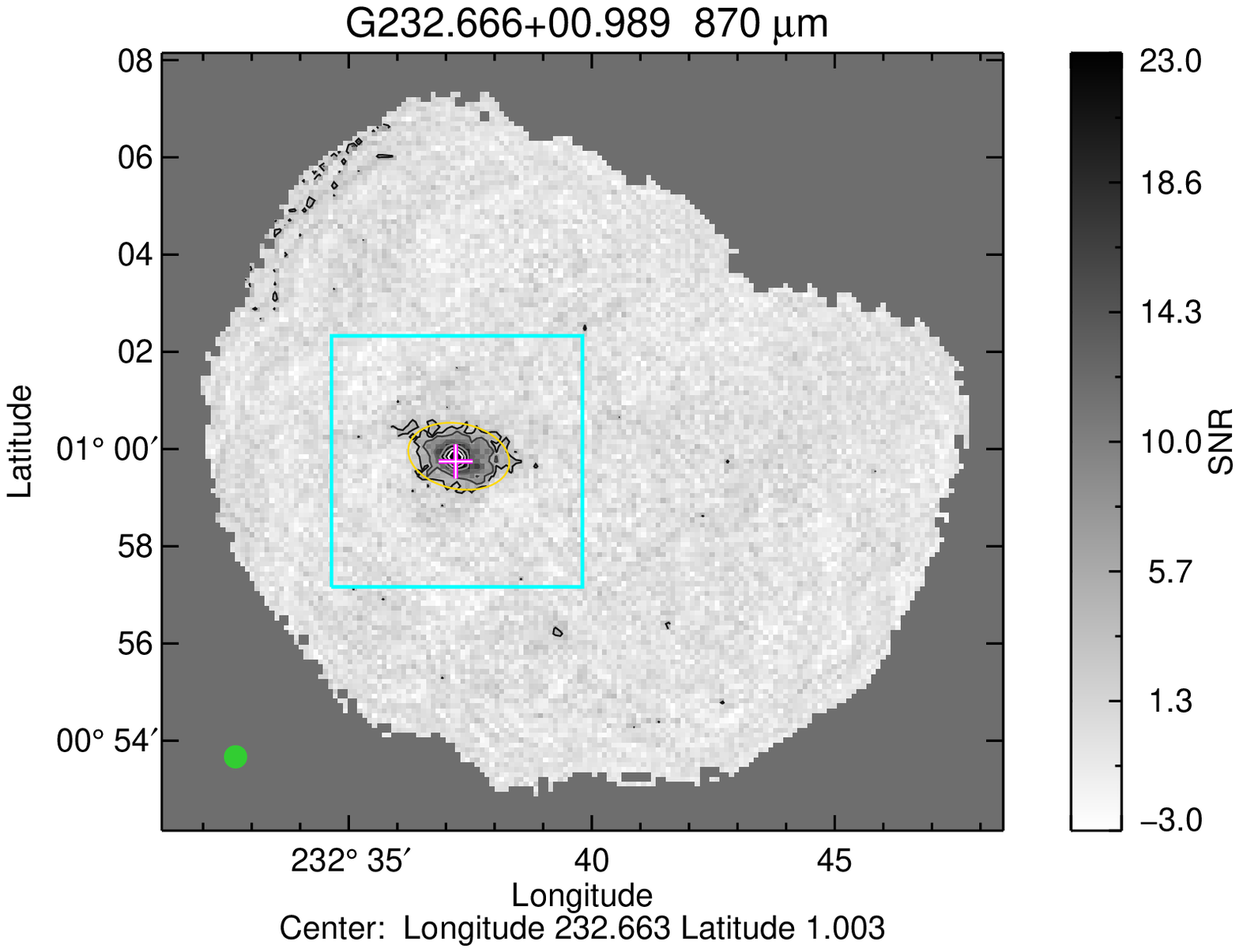}
\includegraphics[width=0.49\textwidth, trim= 0 0 0 0]{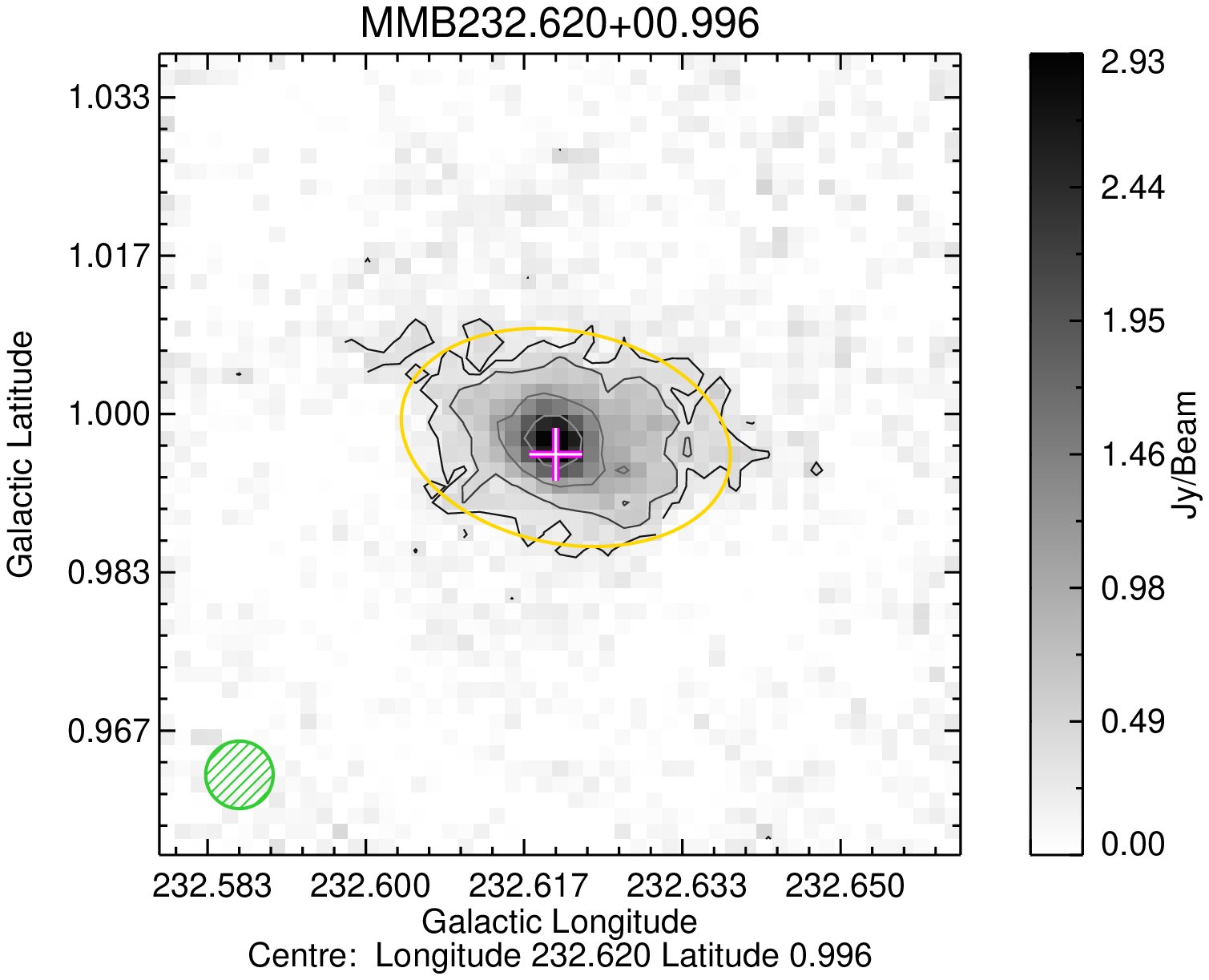}
\includegraphics[width=0.49\textwidth, trim= 0 0 0 0]{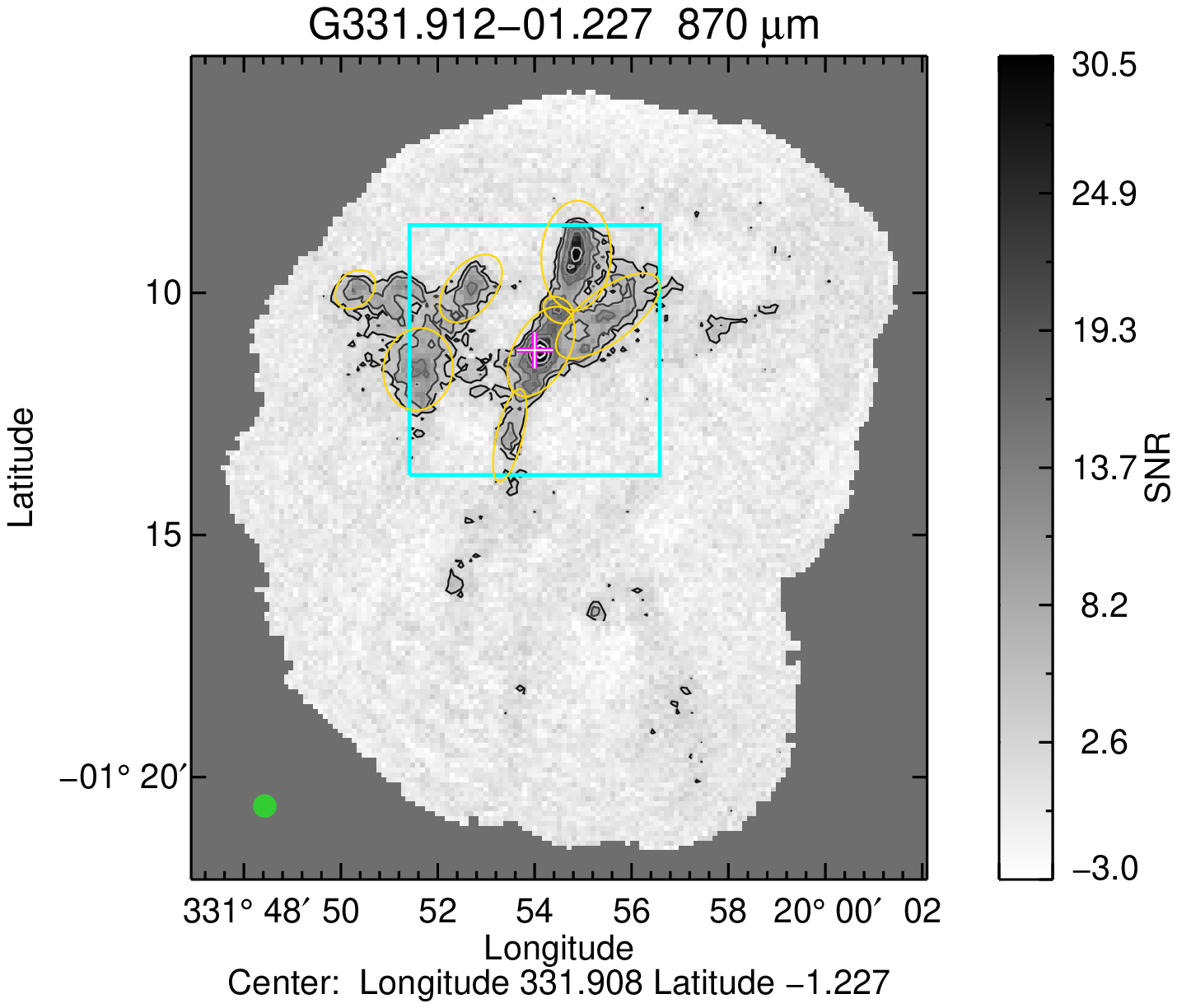}
\includegraphics[width=0.49\textwidth, trim= 0 0 0 0]{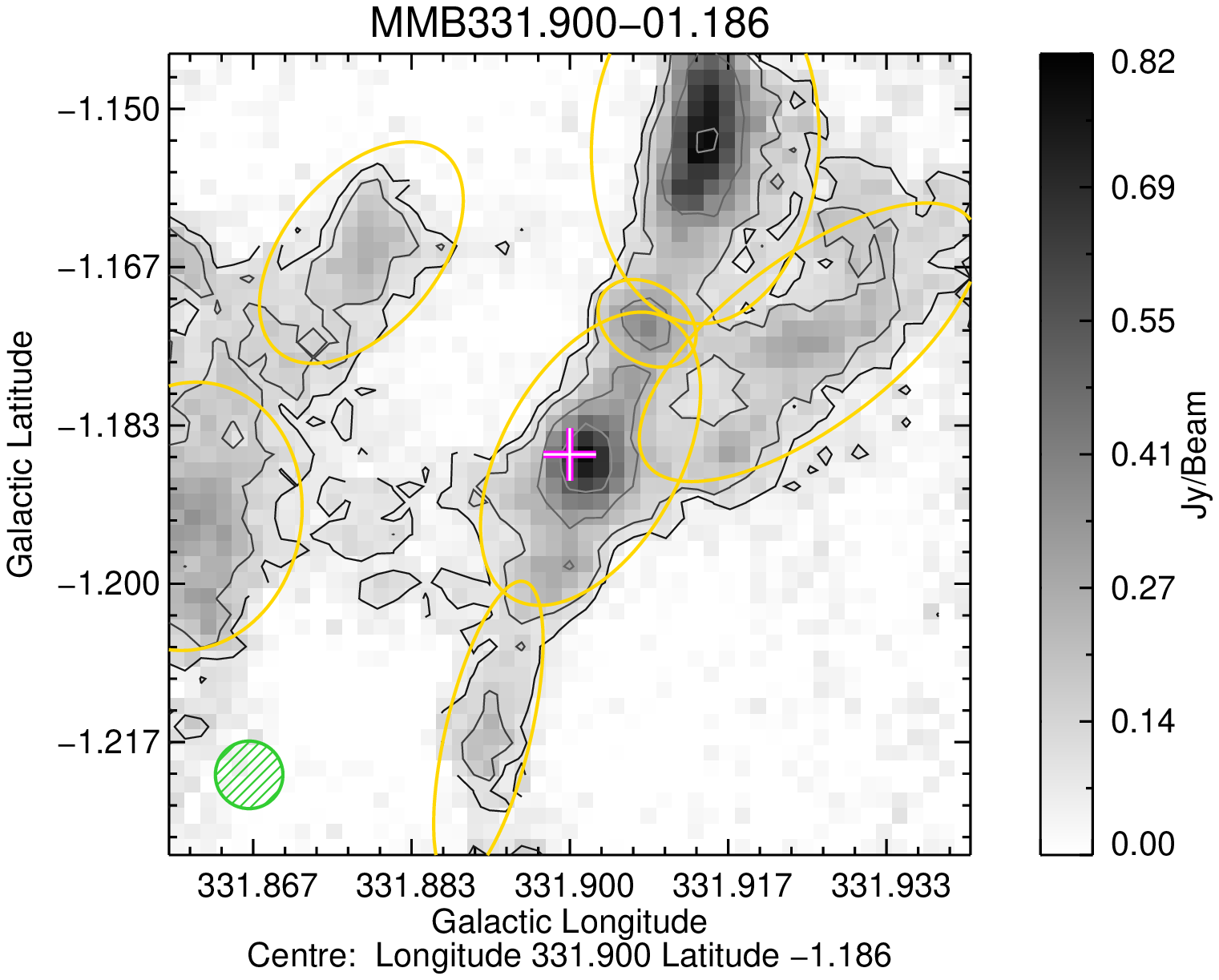}
\includegraphics[width=0.49\textwidth, trim= 0 0 0 0]{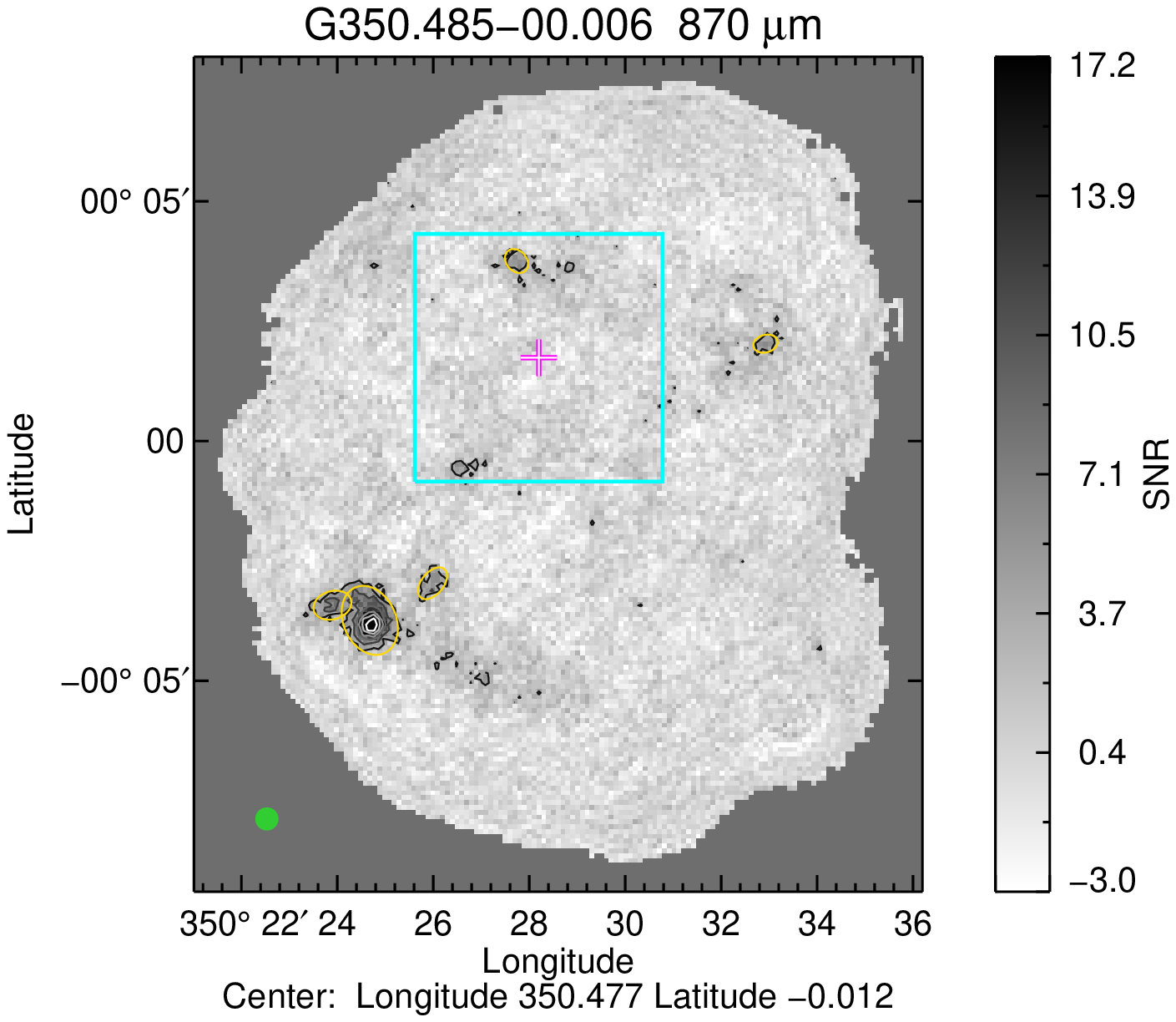}
\includegraphics[width=0.49\textwidth, trim= 0 0 0 0]{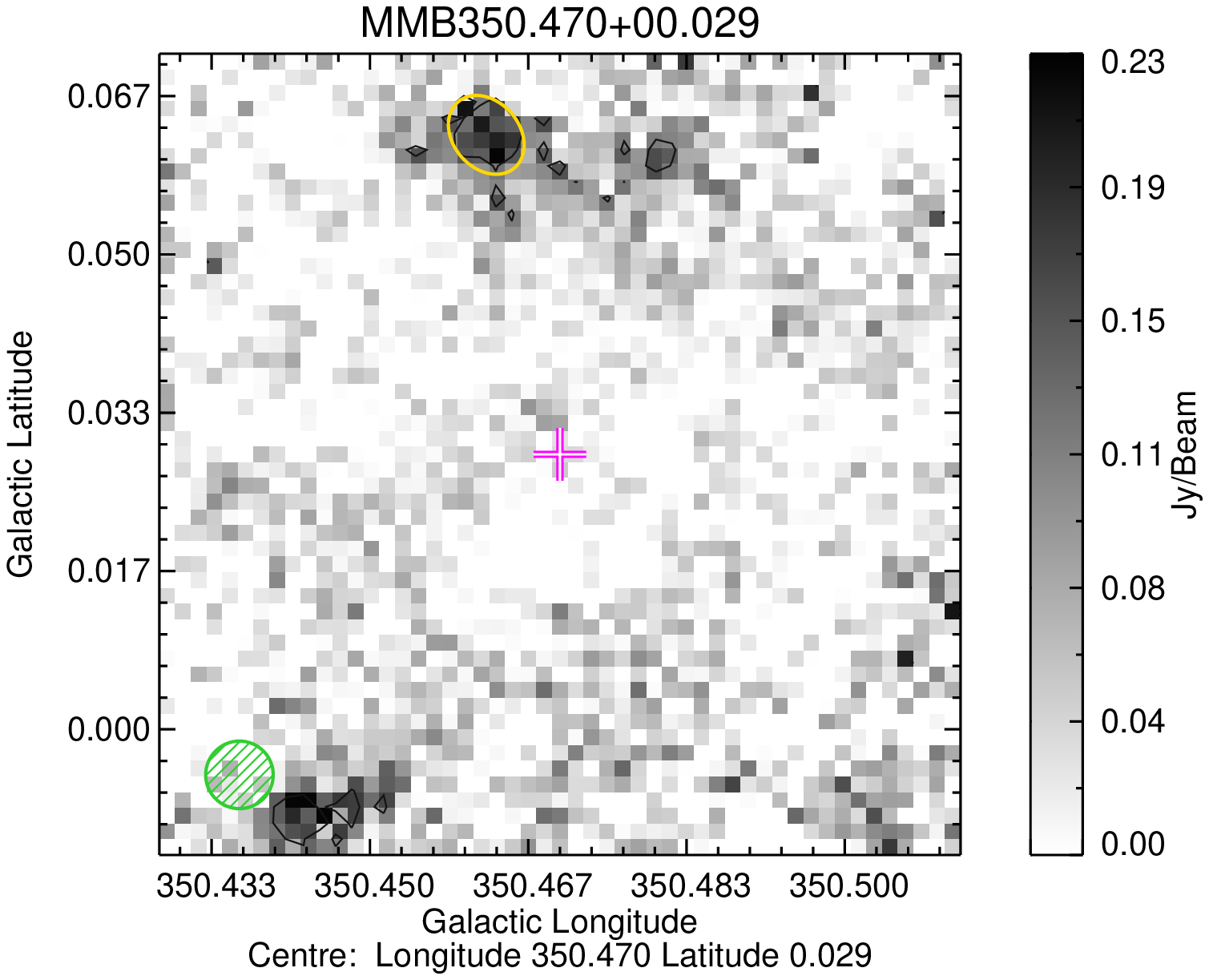}

\caption{\label{fig:irac_images_uchiis}  Examples of the 870-\mum\ continuum emission maps obtained. The left-hand panels show the whole field and the right-hand panels contain smaller maps centred on the position of the methanol maser and outlined in the images of the full fields. The grey scale to the right of each plot shows the intensity and positions of the methanol masers are shown as magenta crosses. The grey contour levels start at 3$\sigma$ and increase in steps set by a dynamically determined power-law of the form $D=3\times N^i+2$, where $D$ is the dynamic range of the \submm\ emission map (defined as the peak brightness divided by the local r.m.s. noise), $N$ is the number of contours used (6 in this case), and $i$ is the contour power-law index (see \citealt{thompson2006} for more details). The lowest power-law index used was one, which results linearly spaced contours starting at 2$\sigma$ and increasing in steps of 3$\sigma$.  The yellow ellipses show the size and orientation of the compact sources identified by \sex. The angular resolution of the APEX beam at 870\,\mum\ is shown by the green circle in the lower left corner of each plot. The full version of this figure is available as supplementary material in the online version of the journal.}

\end{center}
\end{figure*}

We have matched 70 methanol masers with 68 \submm\ sources, two clumps being each associated with two masers. In the upper and middle panels of Fig.\,\ref{fig:irac_images_uchiis} we present maps of the 870-\mum\ emission towards two of these. In Paper\,I there is a strong correlation between the peak \submm\ emission and the position of the methanol maser and this is also the case for these new associations. The median angular offset between the methanol maser position and the peak of the \submm\ emission is $\sim$6\arcsec.

\setlength{\tabcolsep}{6pt}

\begin{table*}

  \begin{center}\caption{\label{tbl:ATLASGAL_dark_sources} MMB sources without a \submm\ counterpart. The distance have been determined from the \citet{brand1993} rotation model and the kinematic distance solutions have been taken from \citet{green2011b}. }
\begin{minipage}{\linewidth}
\begin{tabular}{l.....}
  \hline \hline
    \multicolumn{1}{l}{MMB Name$^{\rm{a}}$}
  &  \multicolumn{1}{c}{MMB Peak Flux} &\multicolumn{1}{c}{870\,\mum\  Upper Limit} &  \multicolumn{1}{c}{Distance} &\multicolumn{1}{c}{$z$}&\multicolumn{1}{c}{Mass Upper Limit} \\
     \multicolumn{1}{c}{}
  &  \multicolumn{1}{c}{(Jy)} &\multicolumn{1}{c}{(Jy\,beam$^{-1}$)} &  \multicolumn{1}{c}{(kpc)} &\multicolumn{1}{c}{(pc)}&\multicolumn{1}{c}{(\msun\,beam$^{-1}$)} \\

  \hline
MMB005.677$-$00.027	&	0.79	&	0.12	&	\multicolumn{1}{c}{$\cdots$}	&	\multicolumn{1}{c}{$\cdots$}	&	\multicolumn{1}{c}{$\cdots$}	\\
MMB014.521+00.155$\star$	&	1.40	&	0.12	&	5.5	&	14.9	&	19.3	\\
MMB293.723$-$01.742	&	0.55	&	0.23	&	9.5	&	-288.3	&	112.9	\\
MMB303.869+00.194	&	0.90	&	0.12	&	3.7	&	12.6	&	9.2	\\
MMB337.517$-$00.348	&	1.50	&	0.10	&	17.1	&	-103.9	&	154.3	\\
MMB350.470+00.029	&	1.44	&	0.08	&	1.2	&	0.6	&	0.6	\\
MMB356.054$-$00.095$\star$	&	0.52	&	0.14	&	\multicolumn{1}{c}{$\cdots$}	&	\multicolumn{1}{c}{$\cdots$}	&	\multicolumn{1}{c}{$\cdots$}	\\
\hline
\end{tabular}\\
$^{\rm{a}}$ Sources that appear to be associated with weak diffuse emission are identified by $\star$.\\ 

\end{minipage}
\end{center}
\end{table*}

\setlength{\tabcolsep}{3pt} 

We have not detected any significant \submm\ emission towards 7 maser sources. However, we note the presence of some diffuse emission towards two of these (MMB014.521+00.155 and MMB356.054$-$00.095; we will take a more detailed look at all of these non-detections in Sect.\,5.2). In the lower panel of Fig.\,\ref{fig:irac_images_uchiis} we present  870-\mum\  emission maps for one of the seven methanol masers towards which no emission is detected. We estimate a 3-$\sigma$ upper limit for the dust emission associated with these masers from the standard deviation of pixels in a box centred on the maser position.  These values are presented in Table\,\ref{tbl:ATLASGAL_dark_sources}. 

The more sensitive observations presented here have resulted in 870-\mum\  emission being detected towards 35 of the methanol masers not previously associated with dust emission in Paper\,I. These additional observations have therefore been $\sim$90\,per\,cent successful in identifying the clumps hosting these methanol masers as well as allowing more stringent upper limits to be placed on the seven unassociated methanol masers. 

\begin{figure}
\begin{center}
\includegraphics[width=0.49\textwidth, trim= 0 0 0 0]{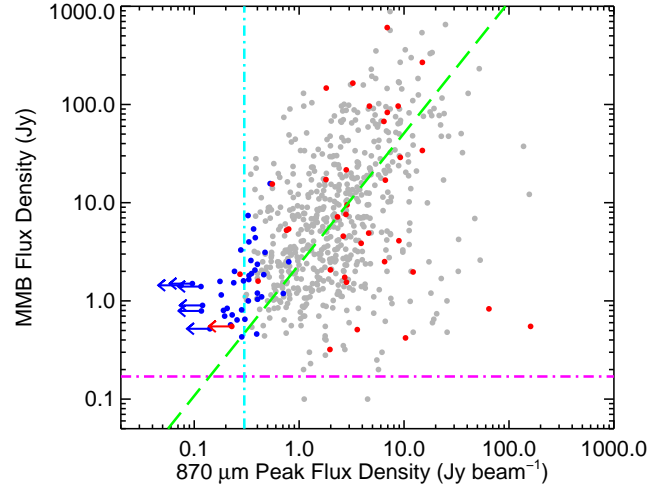}

\caption{\label{fig:flux_density_mmb_atlasgal} Comparison of the methanol-maser flux density and the peak 870-\mum\ flux density of the associated \submm\ clump. Matches with ATLASGAL sources discussed in Paper\,I are shown in grey, sources located outside the ATLASGAL region as shown in red and the deeper integrations are in blue. Upper limits for the non-detections are indicated by a left-facing arrow. The dash-dotted magenta and cyan lines indicate the 5-$\sigma$ sensitivities of the ATLASGAL and MMB surveys, respectively. The results of a Spearman rank correlation test returns a coefficient value of 0.39 and a significance of $\ll 0.01$. The long-dashed green line shows the result of a power-law fit to the data, which has a slope of 1.30$\pm$0.05.
}

\end{center}
\end{figure}

Fig.\,\ref{fig:flux_density_mmb_atlasgal} compares the peak flux density of the methanol maser with the peak 870-\mum\ flux density, showing that the maser and dust emission are correlated. The Spearman correlation coefficient ($r$) is 0.39 with a significance ($p$-value) $\le 0.01$. The methanol masers associated with dust that are located outside the ATLASGAL survey region (red circles) have a similar flux distribution to those identified in Paper\,I (grey circles).  The sources located inside the ATLASGAL survey region that have been re-observed (blue circles) are clearly clustered around the 5$\sigma$ ATLASGAL survey's sensitivity limit (0.3-0.5\,Jy\,beam$^{-1}$). These are the weakest \submm\ sources and seem to follow the overall correlation between the methanol maser and 870-\mum\  emission to lower fluxes. 

\section{Distances}

\subsection{Associations with Complexes}

\subsubsection{Gemini and Auriga Regions}

The Gemini and Auriga regions cover a Galactic longitude range of $\ell=170-194\degr$ and $|b|< 10\degr$ and a velocity range of $-30$\,\kms $<$ \vlsr $<$ $+40$\,\kms. 
There are 6 methanol masers located towards these regions with velocities between $-$5.5 and 18\,\kms\ that are therefore likely to be associated. Since kinematic distances towards the Galactic anti-centre are notoriously unreliable we have adopted the distance (2\,kpc) to these regions reported by \citet{kawamura1998}, with the exception of MMB188.784+01.033 and MMB188.946+00.886 (S\,252) for which maser parallax distances are available (i.e., \citealt{niinuma2011} and \citealt{reid2009b}, respectively). 

\subsubsection{Orion-Monoceros Complex}

This complex covers a Galactic longitude range of $\ell=204\degr-218\degr$ and  $-21\degr < b < -10\degr$ and consists of three giant molecular clouds (GMCs), Orion\,A, Orion\,B and Mon\,R2 (\citealt{wilson2005}). Four methanol masers are found towards these GMCs (\citealt{green2012}). The distance to the Orion complex has been determined from maser parallax measurements to be 414$\pm$7\,pc (\citealt{menten2007}) and this distance has been adopted for all four methanol masers associated with this complex.

\subsubsection{Vela Molecular Ridge}

The Vela Molecular Ridge covers a Galactic longitude range of $\ell$ = 260\degr\ to 270\degr\ and is dominated by nearby emission located at a distance of $\sim$700$\pm200$\,pc (\citealt{netterfield2009}). The complex spans a velocity between 0 and 12\,\kms\ (\citealt{murphy1991,may1988}) and we find 7 of the 8 methanol masers found towards Vela have velocities within a few \kms\ of this range. These methanol masers are likely to be associated with this region and we therefore adopt the complex distance of 0.7\,kpc for these masers. The only methanol maser located towards this region not associated is MMB269.456$-$01.467, which has a velocity of 56.1\,\kms\ and is likely to be located in the background.

\subsection{Literature distances}

In addition to the complexes mentioned we have searched the literature to assign distances to the remaining methanol masers. Kinematic distances have been determined for 17 of these masers by \citet{green2011b}\footnote{Note that \citet{green2011b} used the Galactic rotation model determined by \citet{reid2009}.  To be consistent with Paper\,I we have converted their distances to the \citet{brand1993} rotation model.}, spectrophotometric distances have been determined for 2 masers by \citet{moises2011} and maser parallax distances are available for another 2 sources (MMB196.454-01.677 and MMB232.620+00.996; \citealt{honma2007} and \citealt{reid2009}, respectively). Combined with the sources associated with complexes, these two steps provide distances for 38 of the 77 methanol masers of interest for this study.

\begin{figure}
\begin{center}
\includegraphics[width=0.49\textwidth, trim= 0 0 0 0]{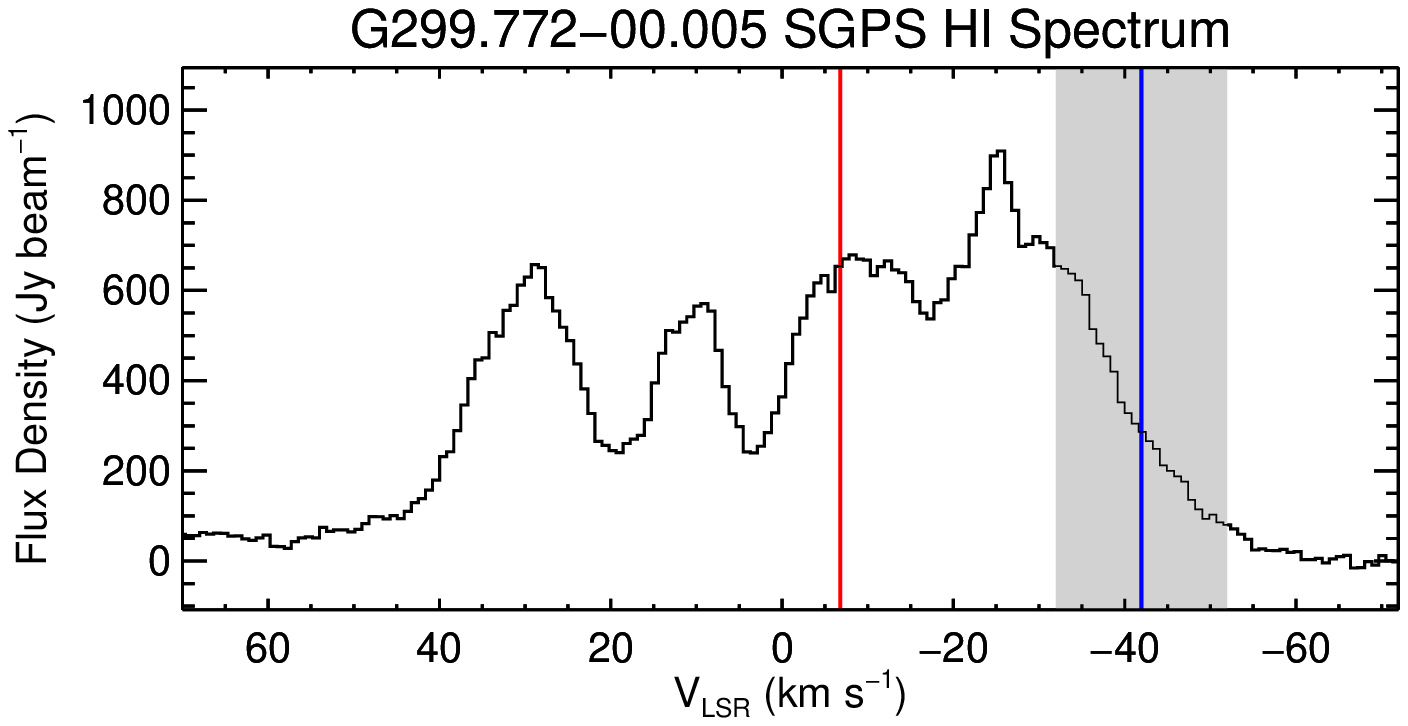}
\includegraphics[width=0.49\textwidth, trim= 0 0 0 0]{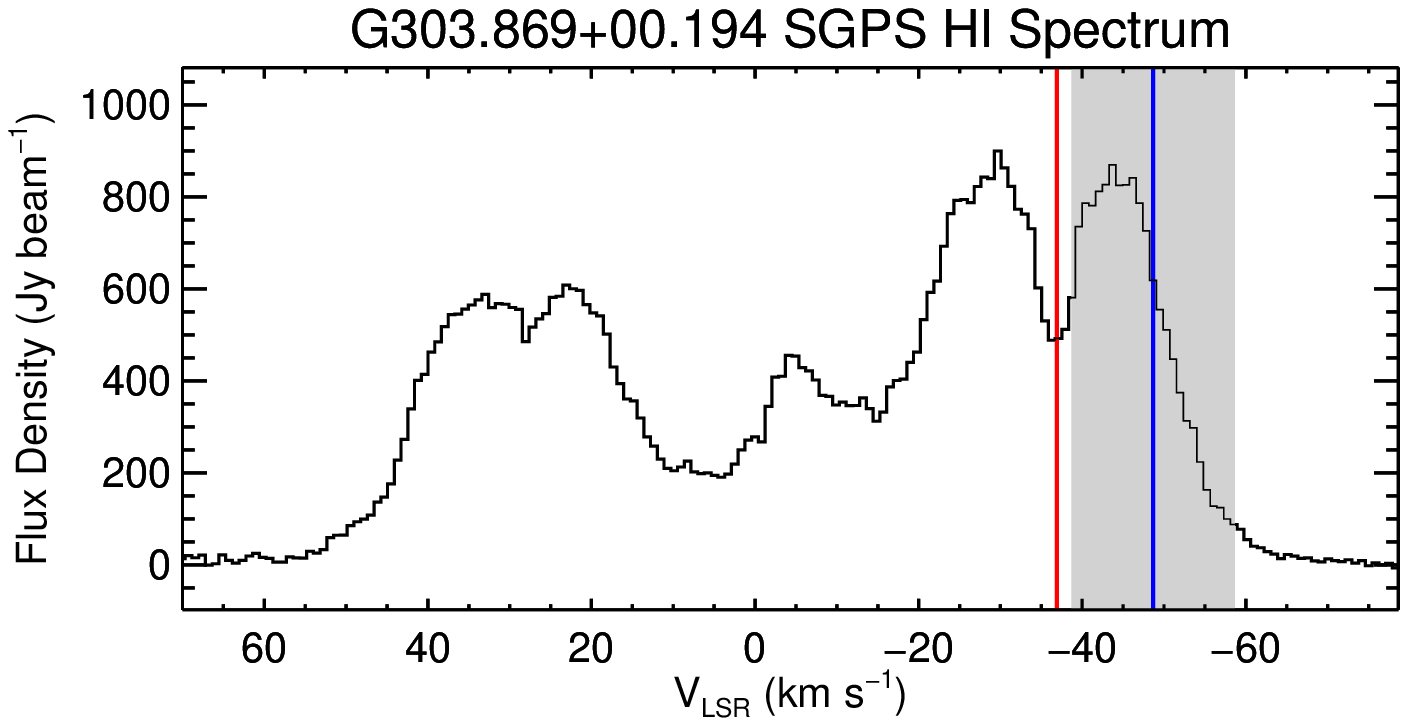}

\caption{\label{fig:hi_spectra} Example SGPS
  \hi\ spectra towards the methanol masers located within the Solar circle for which no reliable distance was available in the literature. The source velocity ($v_{\rm{s}}$) and the velocity of the tangent point ($v_{\rm{t}}$) are shown by the red and blue vertical lines, respectively. The grey vertical band covers the velocity region 10\,\kms\ either side of the tangent velocity, within which sources are assumed to be at the tangent position. The lower panel shows the only source associated with a strong absorption dip at the same velocity as the peak maser emission, which is therefore likely to be located at the near distance. The full version of this figure is available as supplementary material in the online version of the journal.}

\end{center}
\end{figure}

\subsection{Kinematic distances}

We are left with 39 methanol masers for which a distance is not currently available in the literature. For these sources we have determined kinematic distances using the \citet{brand1993} rotation model and assume that the radial velocity of the molecular clump hosting the maser is similar to the velocity of the peak intensity of the maser emission. We exclude four of these sources that are located within 10\degr\ of the Galactic Centre where the large uncertainties associated with kinematic distances render them unreliable. 

Galactic rotation models, such as that of \citet{brand1993}, provide a unique distance for sources located outside the Solar circle (i.e., Galactocentric radius (R$_{\rm{gc}}$) $>$ 8.5\,kpc; we refer to this as the outer Galaxy). However, for sources with $R_{\rm{gc}} < 8.5$\,kpc (inner Galaxy) the radial velocity corresponds to two possible distances spaced equally on either side of the tangent distance, commonly referred to as the \emph{near} and \emph{far} distances.  This degeneracy is known as the kinematic distance ambiguity (KDA).

Of the 35 sources for which we have determined kinematic distances, 16 are located in the outer Galaxy and are therefore not affect by KDAs. A further 5 sources are found to have velocity within $\sim$10\,\kms\ of the tangent velocity, which has been determined from the termination velocity of the Galactic \hi\ emission (i.e., \citealt{mcclure2007}).  These have been placed at the distance of the tangent point. The remaining 14 sources are located within the Solar circle and are associated with KDAs. We solve 6 of these by assigning an upper limit on their allowed distance from the Galactic mid-plane of 120\,pc (i.e., $\sim$5 times the Galactic scale height of massive stars; \citealt{urquhart2014a}); if a far distance would place the source above this limit a near distance is considered more likely. 

For the remaining 8 sources we use \hi\ spectra extracted from the Southern Galactic Plane Survey (SGPS; \citealt{mcclure2005}) to examine the emission profile near to the maser velocity. If we find an absorption dip in the \hi\ spectrum at the same velocity then a near distance is preferred, otherwise a far distance is considered more likely (for a detailed discussion of this method see \citealt{roman2009}). In Fig.\,\ref{fig:hi_spectra} we present the \hi\ spectra extracted towards the positions of two of these methanol masers. From inspection of these spectra we have placed 7 sources at the far distance and one source at the near distance (i.e., MMB303.869+00.194). 

Combining these kinematic distances with those found in the literature discussed in the previous two subsection, we have collated distances for 73 methanol masers in our sample.  The other four sources have been excluded as they were found to be located towards the Galactic Centre. 

\subsection{Distance distribution}

\begin{figure}
\begin{center}
\includegraphics[width=0.49\textwidth, trim= 0 0 0 0]{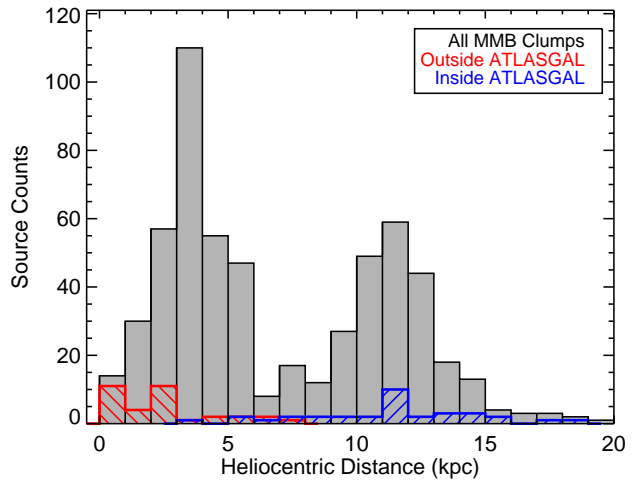}

\caption{\label{fig:distance_hist} Heliocentric distance distribution for all \submm\ clumps associated with a MMB source and for which a distance has been unambiguously determined (grey filled histogram). The new matches are shown in red and blue (see legend for details). The bin size is 1\,kpc.} 

\end{center}
\end{figure}

In addition to the distances for the new detections, we have used new maser parallax measurements (i.e., \citealt{reid2014}) and kinematic solutions derived from \hi\ analysis of ATLASGAL clumps (i.e., Wienen et al., 2014) to refine and expand the number of masers with distance estimates. Furthermore, analysis presented in Paper\,I revealed 4 methanol masers that are associated with clumps having abnormally low masses ($<$1\,\msun) that were clearly distinct from the general distribution; these are MMB348.195+00.768, MMB342.954$-$00.019,  MMB316.412$-$00.308 and MMB344.581$-$00.024. Near distance solutions where found for all of these sources (\citealt{green2011b,urquhart2014b}), albeit with low confidence. It is worth noting that all of these sources have velocities close to zero and therefore their kinematic distances are likely to be very uncertain. Given the large difference between the masses of these associated clumps a near distance is rather unlikely and therefore we have assume far distance for all of these sources. A full discussion of these distance adjustments is beyond the scope of this paper and will be presented in a future research note, however, these distance adjustments do not significantly change the overall distributions of any of the parameters discussed here.

In Fig.\,\ref{fig:distance_hist} we present an updated version of the distance distribution presented in Paper\,I, incorporating the 68 new maser-associated clumps. This plot shows the distance distribution of 540 methanol-maser associated clumps. The overall distribution is bimodal, with peaks at 3-4\,kpc and 11-12\,kpc. The new maser-associated clumps located within the ATLASGAL region are clearly more distant with a peak at $\sim$11-12\,kpc, which is consistent with the dust emission associated with these methanol masers falling below the ATLASGAL survey's sensitivity limit and resulting in their non-detection in Paper I.

The methanol masers that lie outside the ATLASGAL region are either located in the outer Galaxy ($186\degr < \ell < 280\degr$) or are located at high latitudes ($|b| > 1.5\degr$).  In both cases, this subsample is dominated by nearby lower-luminosity objects with only a few sources having distances larger than a few kpc. The two minor peaks seen in the distribution at 0.5 and 2.5\,kpc are the masers associated with the Orion-Monoceros Complex and the Vela Molecular ridge, and the Galactic anti-centre, respectively.

\section{Physical parameters}

\begin{figure*}
\begin{center}

\includegraphics[width=0.45\textwidth, trim= 0 0 0 0]{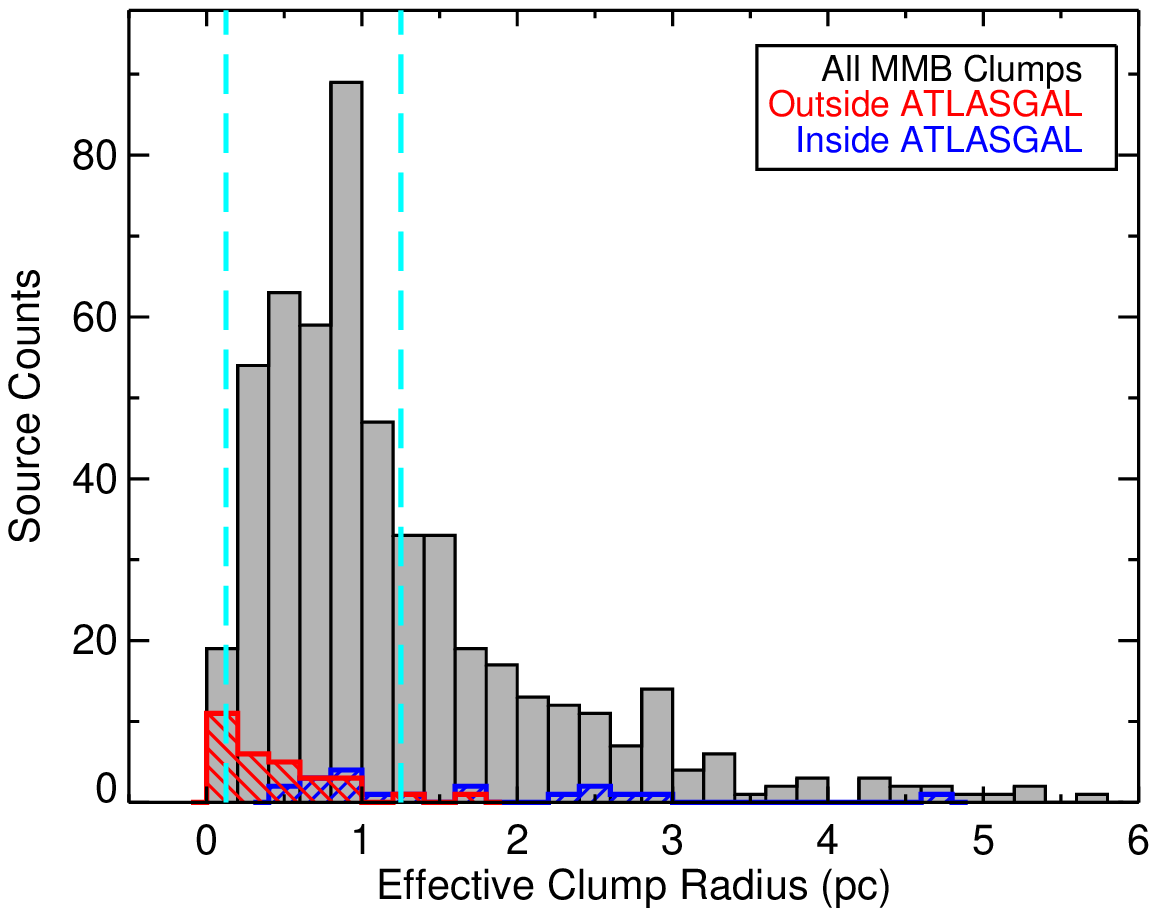}
\includegraphics[width=0.45\textwidth, trim= 0 0 0 0]{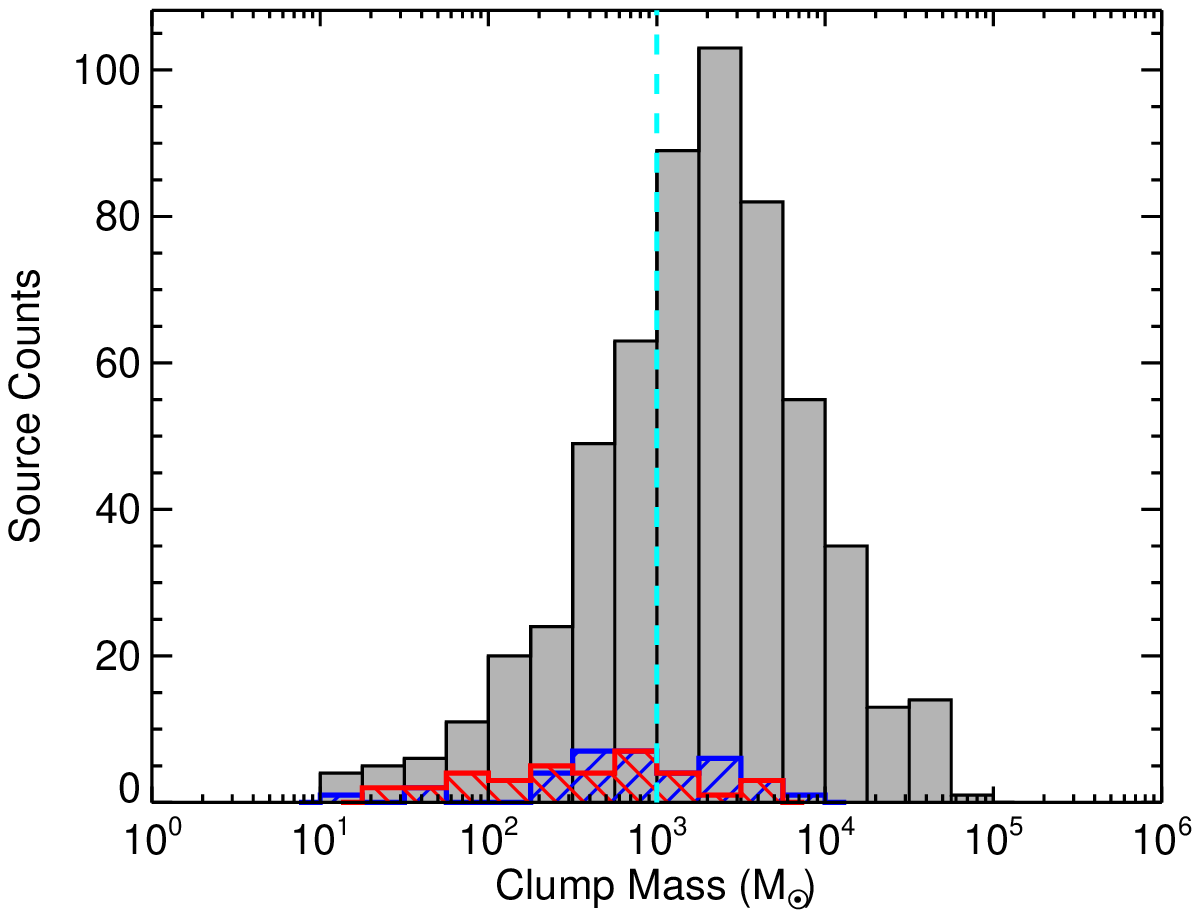}
\includegraphics[width=0.45\textwidth, trim= 0 0 0 0]{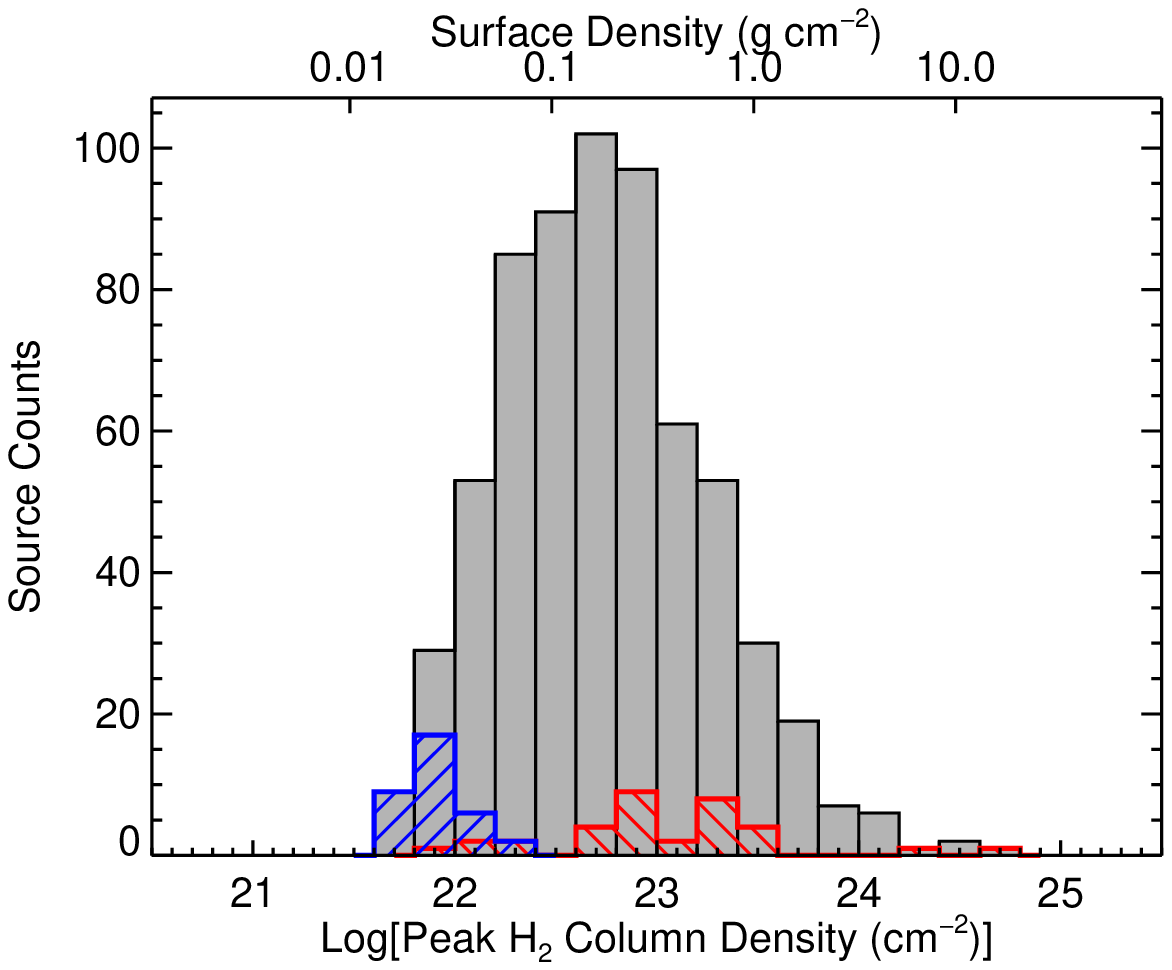}
\includegraphics[width=0.45\textwidth, trim= 0 0 0 0]{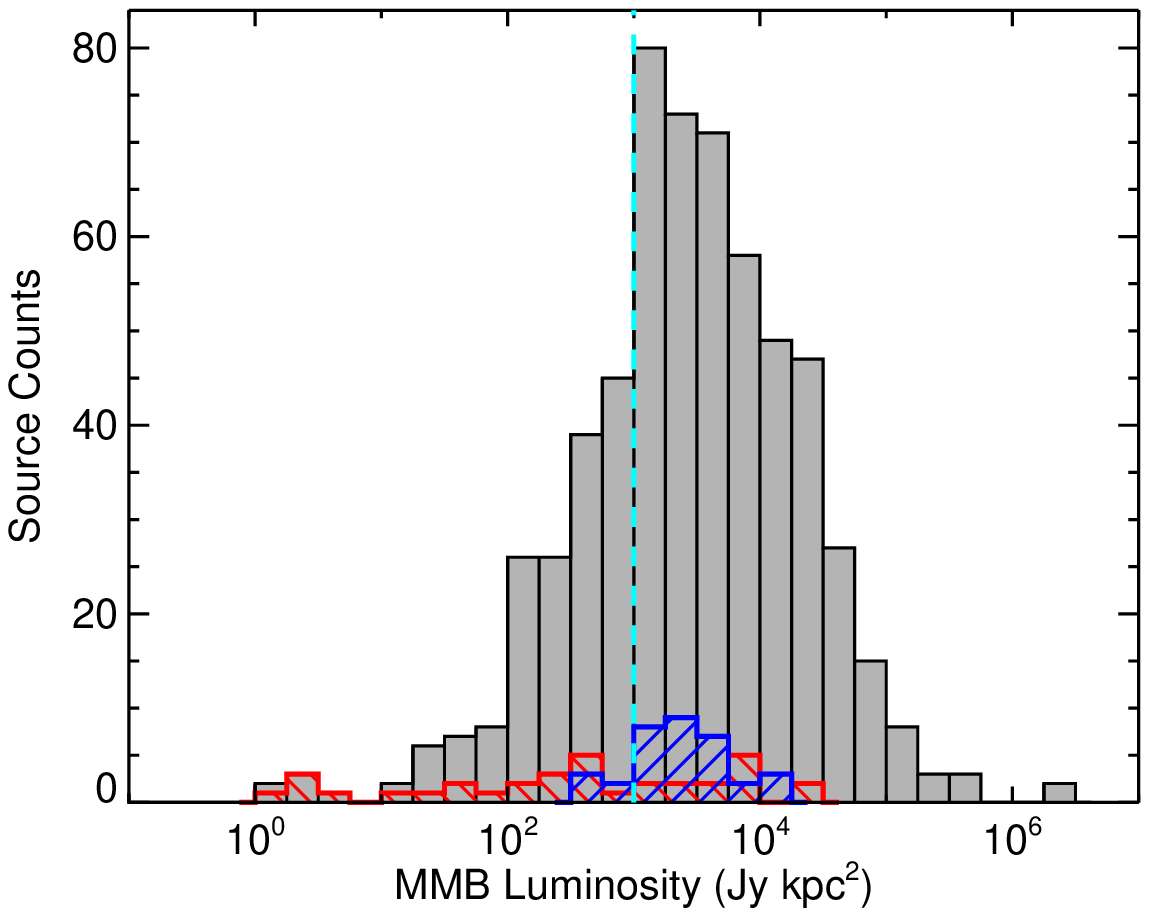}

\caption{\label{fig:mass_distribution} Distributions of derived parameters. Upper left panel: Effective radius: the left and right dashed vertical lines in the upper panel indicate the radii separating cores and clumps (0.125\,pc), and clumps and clouds (1.25\,pc), respectively (i.e., \citealt{bergin2007}). Upper right panel: The isothermal dust masses assuming a dust temperature of 20\,K. The vertical blue line indicates the completeness limit (see Paper\,I for details). Lower left panel: Column density. Lower right panel: Methanol-maser luminosity.}

\end{center}

\end{figure*}

\setlength{\tabcolsep}{3pt}

\begin{table*}

\begin{center}\caption{Derived clump parameters. The columns are as follows: (1) and (2) MMB and \submm\ source names; (3) angular offset between the peak of the submillimetre emission and the maser position; (4) ratio of the semi-major to semi-minor sizes of the ATLASGAL source; (5) ratio of the integrated and peak submillimetre emission ($Y$-factor); (6) heliocentric distance; (7) heliocentric distance reference; (8) Galactocentric distance; (9) effective physical radius; (10) column density; (11) clump mass derived from the integrated 870-\mum\ emission assuming a dust temperature of 20\,K; (12) isotropic methanol maser luminosity. }
\label{tbl:derived_clump_para}
\begin{minipage}{\linewidth}
\small
\begin{tabular}{ll....c.....}
\hline \hline
   \multicolumn{1}{c}{MMB name}&  \multicolumn{1}{c}{Submm name$^{\rm{a}}$}&	\multicolumn{1}{c}{Offset}  &\multicolumn{1}{c}{Aspect}  &\multicolumn{1}{c}{$Y$-factor}  &	\multicolumn{1}{c}{Distance} &	\multicolumn{1}{c}{Distance}&\multicolumn{1}{c}{$R_{\rm{GC}}$}&\multicolumn{1}{c}{Radius}&\multicolumn{1}{c}{Log($N$(H$_2$))} &	\multicolumn{1}{c}{Log($M_{\rm{Clump}}$)}&	\multicolumn{1}{c}{Log($L_{\rm{MMB}}$)}\\
    \multicolumn{1}{c}{ }&  \multicolumn{1}{c}{ }&	\multicolumn{1}{c}{($^{\prime\prime}$)}  &\multicolumn{1}{c}{Ratio}  &\multicolumn{1}{c}{ }  &	\multicolumn{1}{c}{(kpc)}&\multicolumn{1}{c}{Ref.}  &\multicolumn{1}{c}{(kpc)}&\multicolumn{1}{c}{(pc)}&\multicolumn{1}{c}{(cm$^{-2}$)} &	\multicolumn{1}{c}{(\msun)}&	\multicolumn{1}{c}{(Jy\,kpc$^2$)}\\
        \multicolumn{1}{c}{(1)}&  \multicolumn{1}{c}{(2)}&	\multicolumn{1}{c}{(3)}  &\multicolumn{1}{c}{(4)}  &\multicolumn{1}{c}{(5)}  &	\multicolumn{1}{c}{(6)} &\multicolumn{1}{c}{(7)}&\multicolumn{1}{c}{(8)}&\multicolumn{1}{c}{(9)} &	\multicolumn{1}{c}{(10)}&	\multicolumn{1}{c}{(11)}&	\multicolumn{1}{c}{(12)}\\

\hline
MMB006.881+00.093	&	G006.881+00.093	&	0.0	&	1.5	&	4.17	&	 \multicolumn{1}{c}{$\cdots$}	&	\multicolumn{1}{c}{$\cdots$}	&	 \multicolumn{1}{c}{$\cdots$}	&	 \multicolumn{1}{c}{$\cdots$}	&	22.07	&	 \multicolumn{1}{c}{$\cdots$}	&	 \multicolumn{1}{c}{$\cdots$}	\\
MMB012.776+00.128	&	G012.776+00.128	&	0.0	&	1.2	&	1.30	&	13.0	&	1	&	5.0	&	 \multicolumn{1}{c}{$\cdots$}	&	21.71	&	2.40	&	3.25	\\
MMB013.696$-$00.156	&	G013.691$-$00.158	&	19.2	&	1.2	&	4.61	&	10.9	&	1	&	3.3	&	1.75	&	21.95	&	3.03	&	3.45	\\
MMB015.607$-$00.255	&	G015.605$-$00.255	&	6.0	&	3.9	&	12.30	&	11.4	&	1	&	4.0	&	0.93	&	21.85	&	3.40	&	2.85	\\
MMB016.864$-$02.159	&	G016.866$-$02.159	&	6.2	&	1.5	&	3.14	&	1.7	&	2	&	6.9	&	0.19	&	23.36	&	2.64	&	3.00	\\
MMB016.976$-$00.005	&	G016.976$-$00.003	&	6.1	&	1.2	&	3.41	&	15.5	&	1	&	7.8	&	 \multicolumn{1}{c}{$\cdots$}	&	21.81	&	3.06	&	3.29	\\
MMB017.021$-$02.403	&	G017.021$-$02.403	&	0.1	&	1.1	&	3.97	&	2.4	&	2	&	6.2	&	0.42	&	23.06	&	2.77	&	2.55	\\
MMB018.341+01.768	&	G018.341+01.768	&	0.3	&	1.7	&	5.74	&	2.6	&	2	&	6.1	&	0.65	&	23.07	&	3.01	&	3.92	\\
MMB018.440+00.045	&	G018.440+00.036	&	30.4	&	2.1	&	11.51	&	11.6	&	1	&	4.4	&	2.60	&	22.06	&	3.59	&	3.50	\\
MMB019.614+00.011	&	G019.612+00.011	&	6.0	&	1.2	&	3.87	&	13.1	&	1	&	5.9	&	0.45	&	21.92	&	3.09	&	3.94	\\
\hline\\
\end{tabular}\\
Distance References: (1) \citet{green2011b}; (2) this paper; (3) \citet{niinuma2011}; (4) \citet{oh2010}; (5) \citet{kawamura1998}; (6) \citet{honma2007}; (7) \citet{menten2007}; (8) \citet{reid2009}; (9) \citet{netterfield2009}; (10) \citet{moises2011}.
Notes: Only a small portion of the data is provided here, the full table is available in electronic form at the CDS via anonymous ftp to cdsarc.u-strasbg.fr (130.79.125.5) or via http://cdsweb.u-strasbg.fr/cgi-bin/qcat?J/MNRAS/.

\end{minipage}

\end{center}
\end{table*}

\setlength{\tabcolsep}{6pt}

In the next few subsections we provide a brief outline of the steps used to derive the clump physical properties and we refer the reader to Paper\,I for further details. Fig.\,\ref{fig:mass_distribution} shows the distribution of the new matches with respect to that of the full sample. The derived properties of the methanol masers and their host clump are listed in Table\,\ref{tbl:derived_clump_para}.

\subsection{Effective radius}
\label{sect:size}

The source effective radius is calculated from the geometric mean of the beam-deconvolved major and minor axes (Eqn.\,6 of \citealt{rosolowsky2010}):

\begin{equation}
\label{radius}
\theta_{R_{\rm{eff}}}= \eta \left[(\sigma_{\rm{maj}}^2-\sigma_{\rm{bm}}^2)(\sigma_{\rm{min}}^2-\sigma_{\rm{bm}}^2)\right]^{1/4},
\end{equation}

\noindent where $\sigma_{bm}$ is the \rms\ size of the beam (i.e.,
$\sigma_{bm}=\theta_{\mathrm{FWHM}}/\sqrt{8\ln 2} \simeq 8''$). Following \citet{rosolowsky2010} to be consistent with the analysis presented in Paper\,I we adopt a value for $\eta$ of 2.4; this is roughly equivalent to the usual source effective radius $R_{\rm{eff}}=\sqrt(A/\pi)$ (where $A$ is the surface area of the source; \citealt{dunham2011}). Combining this angular size with the kinematic distances we calculate the physical size of the host dust clump. We present the distribution of the clump sizes for the whole sample in the upper left panel of Fig.\,\ref{fig:mass_distribution} and give the sample statistics in Table\,\ref{tbl:derived_para}. Although the sample covers a large range of physical scales, from  sizes expected for individual cores to whole molecular clouds, the majority falls into the size range expected for clumps and for simplicity we will use this terminology to describe the whole sample. 

\subsection{Clump mass and peak column densities}

The clump masses are estimated from the integrated \submm\ flux density using a temperature of 20\,K and assuming that the dust emission is optically thin, using

\begin{equation}
M \, = \, \frac{D^2 \, S_\nu \, R}{B_\nu(T_D) \, \kappa_\nu},
\end{equation}

\noindent where $S_\nu$ is the integrated 870-\mum\ flux, $D$ is the heliocentric distance to the source, $R$ is the gas-to-dust mass ratio (assumed to be 100), $B_\nu$ is the Planck function for a dust temperature $T_D$, and $\kappa_\nu$ is the dust absorption coefficient taken as 1.85\,cm$^2$\,g$^{-1}$ (see Paper\,I and \citet{schuller2009} for more details). Clump peak column densities are estimated from the peak 870-\mum\ flux density, again assuming an isothermal dust temperature of 20\,K, using the following equation:

\begin{equation}
N_{H_2} \, = \, \frac{S_\nu \, R}{B_\nu(T_D) \, \Omega \, \kappa_\nu \, \mu
\, m_H},
\end{equation}

\noindent where $\Omega$ is the beam solid angle, $\mu$ is the mean molecular weight of the molecular interstellar medium, which we assume to be equal to 2.8, and $m_H$ is the mass of an hydrogen atom.

The mass and column density distributions are presented in the upper-right and lower-left panels of Fig.\,\ref{fig:mass_distribution}, respectively. 
The mean clump mass in this sample is approximately 2000\,\msun, which is sufficient to produce a cluster of stars that includes at least one massive star with a stellar mass of $\ge$20\,\msun (assuming a standard initial mass function and 30\,per\,cent star-formation efficiency, e.g., \citealt{lada2003}). 
Also, the peak 870-\mum\ column density of the new maser-associated clumps measured with the more sensitive observations are among the lowest detected.  This may be because these sources are among the most distant. Fig.\,\ref{fig:mass_distribution} clearly illustrates the beam-dilution effect of distance upon measured column density, with sources in both new subsets having similar masses but with column densities separated by an order of magnitude.

\subsection{Methanol maser luminosities}

Integrated fluxes are not yet available for the MMB catalogue and we must therefore use the peak 6.7-GHz  
flux density to obtain an estimate of the methanol maser's ``luminosity'' using the following equation and assuming the emission is isotropic: 

\begin{equation}
L_{\rm{MMB}} \, = \, 4\pi D^2 S_\nu
\end{equation}

\noindent where $D$ is the heliocentric distance in kpc and $S_\nu$ is the methanol maser peak flux density in Jy. $L_{\rm{MMB}}$ thus has units of Jy\,kpc$^2$. It should be noted here that, in general, astronomical masers are not expected to radiate isotropically, especially if they are amplifying a background continuum source.

The luminosity  distribution of the whole dust-associated sample is shown in the upper right panel of Fig.\,\ref{fig:mass_distribution}. We note the presence of a small peak at the lower end of the main distribution, which was not present in the previous plot presented in Paper\,I. This peak includes all of the methanol masers associated with the Orion-Monoceros complex and the Vela Molecular Ridge, which are two of the nearest massive star forming regions. We are therefore likely to be picking up weak methanol masers, perhaps associated with a population of very young and/or relatively low-luminosity protostellar objects, that are largely missed towards other more distant massive star-forming regions.  

\setlength{\tabcolsep}{6pt}

\begin{table*}

\begin{center}\caption{Summary of derived parameters.}
\label{tbl:derived_para}
\begin{minipage}{\linewidth}
\small
\begin{tabular}{lc.......}
\hline \hline
  \multicolumn{1}{l}{Parameter}&  \multicolumn{1}{c}{Number}&	\multicolumn{1}{c}{Mean}  &	\multicolumn{1}{c}{Standard Error} &\multicolumn{1}{c}{Standard Deviation} &	\multicolumn{1}{c}{Median} & \multicolumn{1}{c}{Min}& \multicolumn{1}{c}{Max}\\
\hline
Effective Radius (pc) &          518&1.22&0.04 & 0.96 & 0.93 & 0.04 & 5.63\\
Log[Clump Mass (\msun)] &           574&3.23&0.03 & 0.67 & 3.28 & 
1.02 & 4.99\\
  Log[Column Density (cm$^{-2}$)] &          646&22.74&0.02 & 0.51 & 22.71 & 
21.65 & 24.60\\
Log[$L_{\rm{MMB}}$ (Jy km\,s$^{-1}$\,kpc$^2$)] &          602&3.42&0.04 & 0.90 & 3.46
 & -1.10 & 6.34\\
\hline\\
\end{tabular}\\

\end{minipage}

\end{center}
\end{table*}

\section{Discussion}

\begin{figure}
\begin{center}
\includegraphics[width=0.49\textwidth, trim= 0 0 0 0]{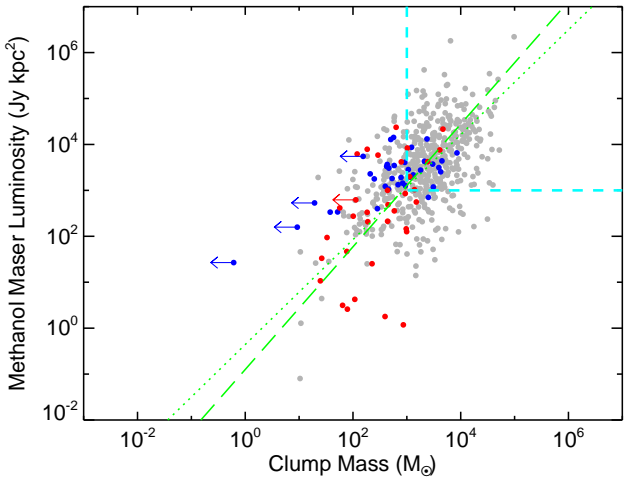}

\caption{\label{fig:maser_mass_plot} The relationship between clump mass and methanol-maser luminosity for the 527 MMB-clump associations for which we have a distance.  Colours are as described in Fig.\,1.  The dashed box shows the region of parameter space within which the sample is complete at a distance of 20\,kpc, containing 312 sources. The long-dashed and dotted green lines show the power-law fits to the whole sample and to a distance-limited subsample (between 3 and 5\,kpc). The slopes agree within the errors and the solution for the distance-limited sample is Log($L_{\rm{MMB}})$=$(1.17\pm0.06) \times {\rm{Log}}(M_{\rm{clump}})+(-0.47\pm0.20)$. A partial-Spearman correlation test (\citealt{collins1998}) to remove the dependency of both of parameters on distance yields a coefficient of 0.41 with a $p$-value $\ll 0.01$.} 

\end{center}
\end{figure}

\subsection{The nature of the new detections}

\begin{figure}
\begin{center}
\includegraphics[width=0.49\textwidth, trim= 0 0 0 0]{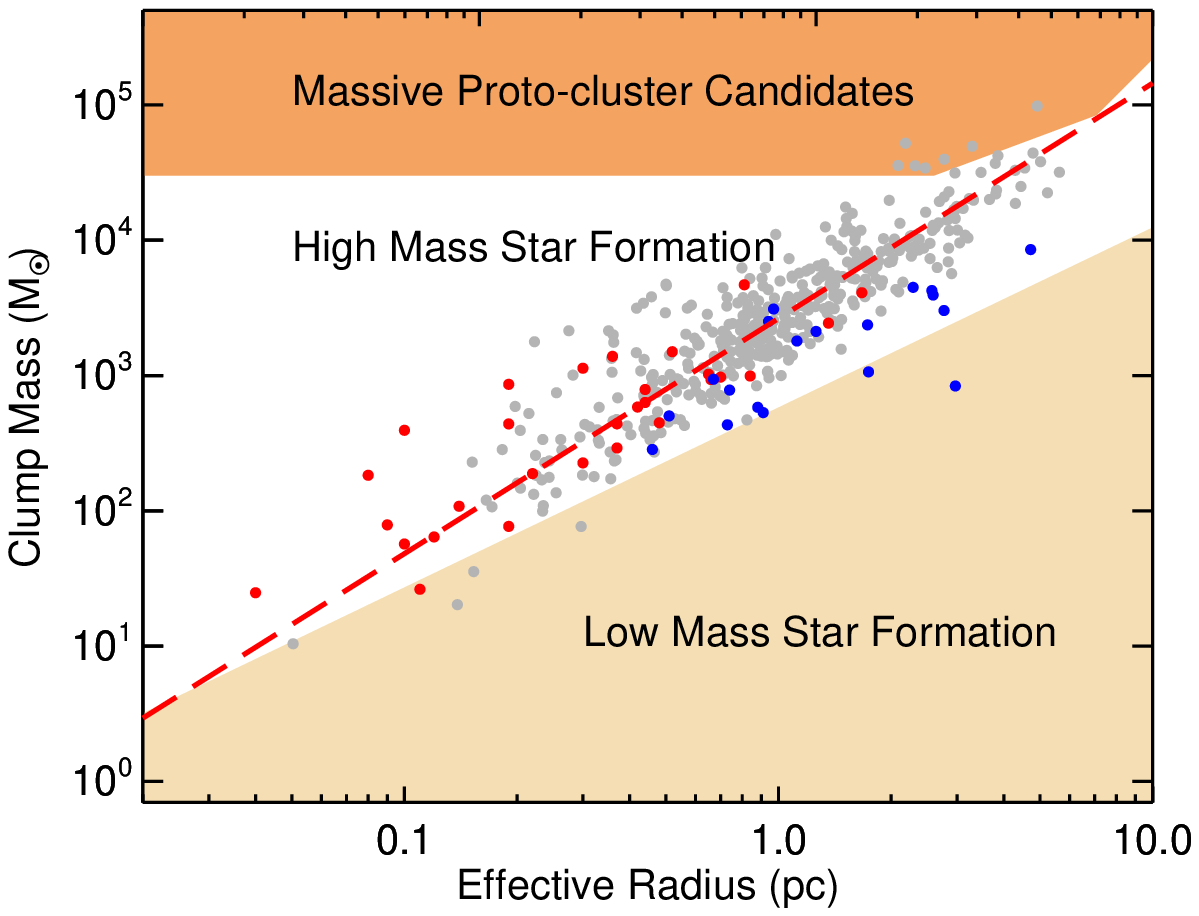}

\caption{\label{fig:mass_radius_distribution} The mass-size relationship of all methanol masers associated with compact \submm\ emission. The beige shaded region shows the area of parameter space found to be devoid of massive star formation, that satisfies the relationship $m(r) \le 580$\,\msun\, $(R_{\rm{eff}}/{\rm{pc}})^{1.33}$ (\citealt{kauffmann2010b}). The orange shaded area towards the top of the diagram indicates the region in which young massive cluster progenitors are expected to be found (\citealt{bressert2012}). A power-law fit to the data gives Log($M_{\rm{clump}})$=$(1.74\pm0.04) \times {\rm{Log}}(R_{\rm{eff}})+(3.42\pm0.01)$. The partial Spearman correlation test yields a correlation coefficient of  0.81 with a $p$-value $\ll$ 0.01, independent of mutual dependence on distance. The sample size is 434.}

\end{center}
\end{figure} 

As mentioned in the Introduction, the analysis of the ATLASGAL and MMB catalogues in Paper I failed to find a match for 41 methanol masers, which led to some speculation as to their nature. The more sensitive observations reported here have resulted in the detection of 870-\mum\  continuum emission towards all but six of these. We have also detected compact emission from cold dust towards 35 of the 36 masers located outside the ATLASGAL survey region. Putting these results in a more global context, when we combine the 70 methanol masers detected here with the observations presented in Paper\,I, we have detections of compact \submm\ emission in 700 sources from the MMB catalogue out of a total of 707, which corresponds to 99\,per\,cent of the whole catalogue. Furthermore, we find the clump masses, maser luminosities and sizes to be similar to those of the larger sample presented and discussed in Paper\,I. We have therefore confirmed that the vast majority of methanol masers are associated with dense clumps and thus with the star-formation process.

Figs.\,\ref{fig:maser_mass_plot} and \ref{fig:mass_radius_distribution} show the mass-luminosity and mass-size relations for all the maser-clump matches presented in the previous paper with the new matches overplotted. These plots are discussed in detail in Paper\,I and so we will not dwell on the significance of these correlations here and we refer the reader to that paper for more information. 

The similarity between the distribution of the new detections in Figs.\,\ref{fig:maser_mass_plot} and  \ref{fig:mass_radius_distribution} and of the slopes derived from power-law fits with those determined in Paper\,I, would suggest that the combined sample represents a single population and that all sources are likely to be tracing the same phenomenon. Furthermore, $\sim$99\,per\,cent of the methanol masers are associated with clumps that satisfy the \citet{kauffmann2010b} criterion for massive star formation (only a handful of methanol-maser associated clumps fail this test). This provides the \emph{strongest evidence so far} that methanol masers are an almost ubiquitous tracer of massive star formation.  

\subsection{Nature of the methanol maser non-detections}

\begin{figure*}
\begin{center}
\includegraphics[width=0.33\textwidth, trim= 0 0 0 0]{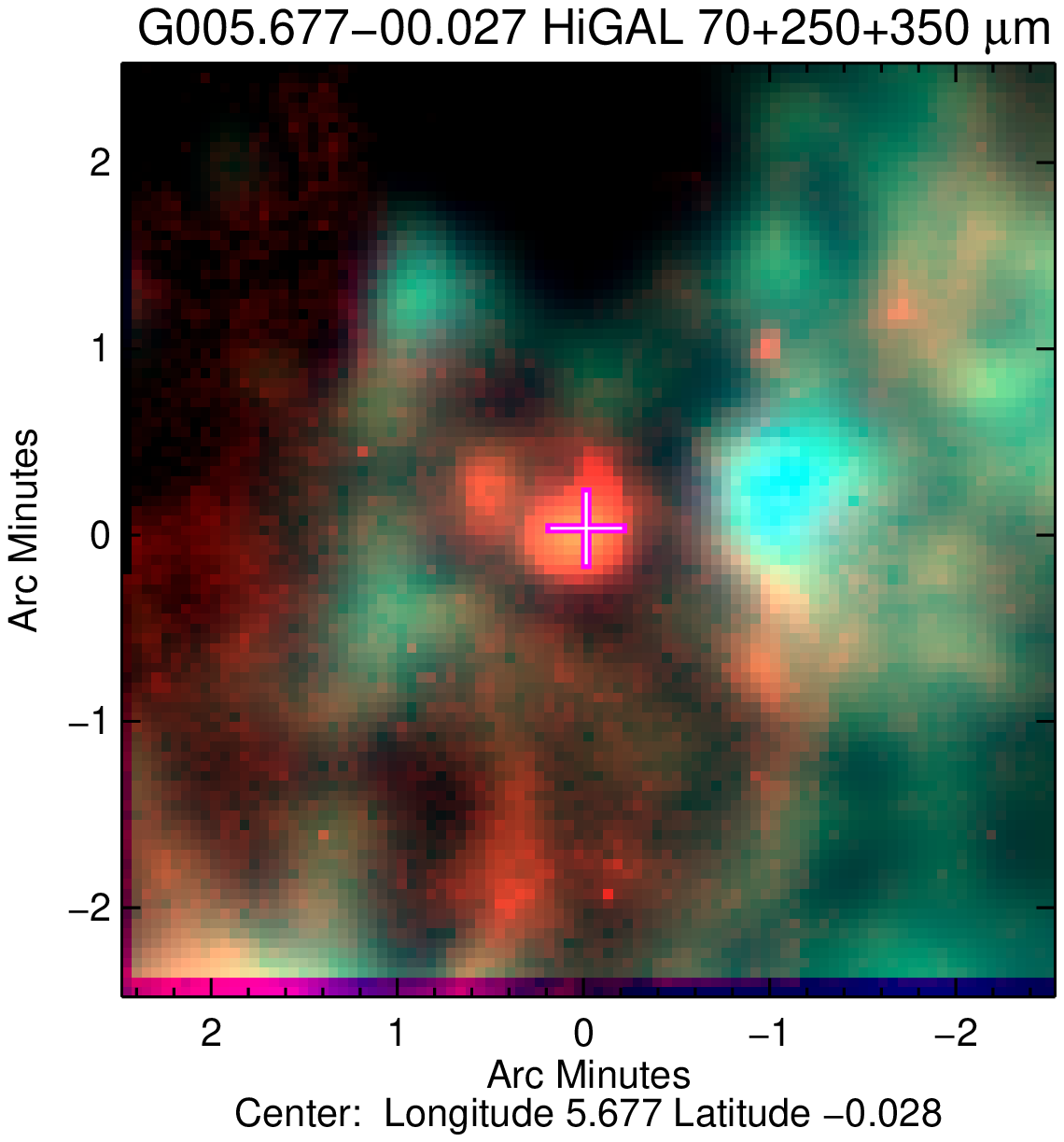}
\includegraphics[width=0.33\textwidth, trim= 0 0 0 0]{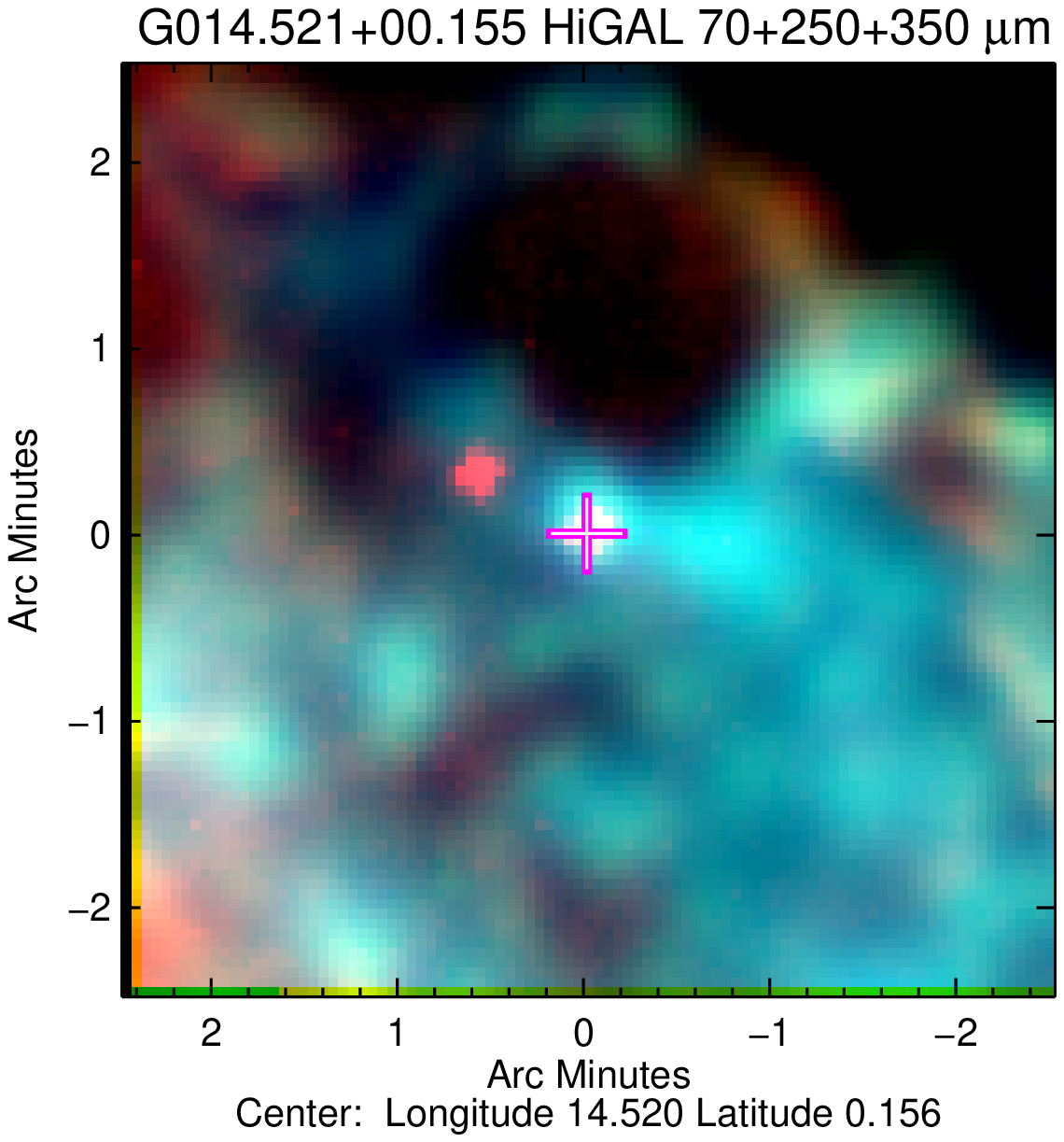}
\includegraphics[width=0.33\textwidth, trim= 0 0 0 0]{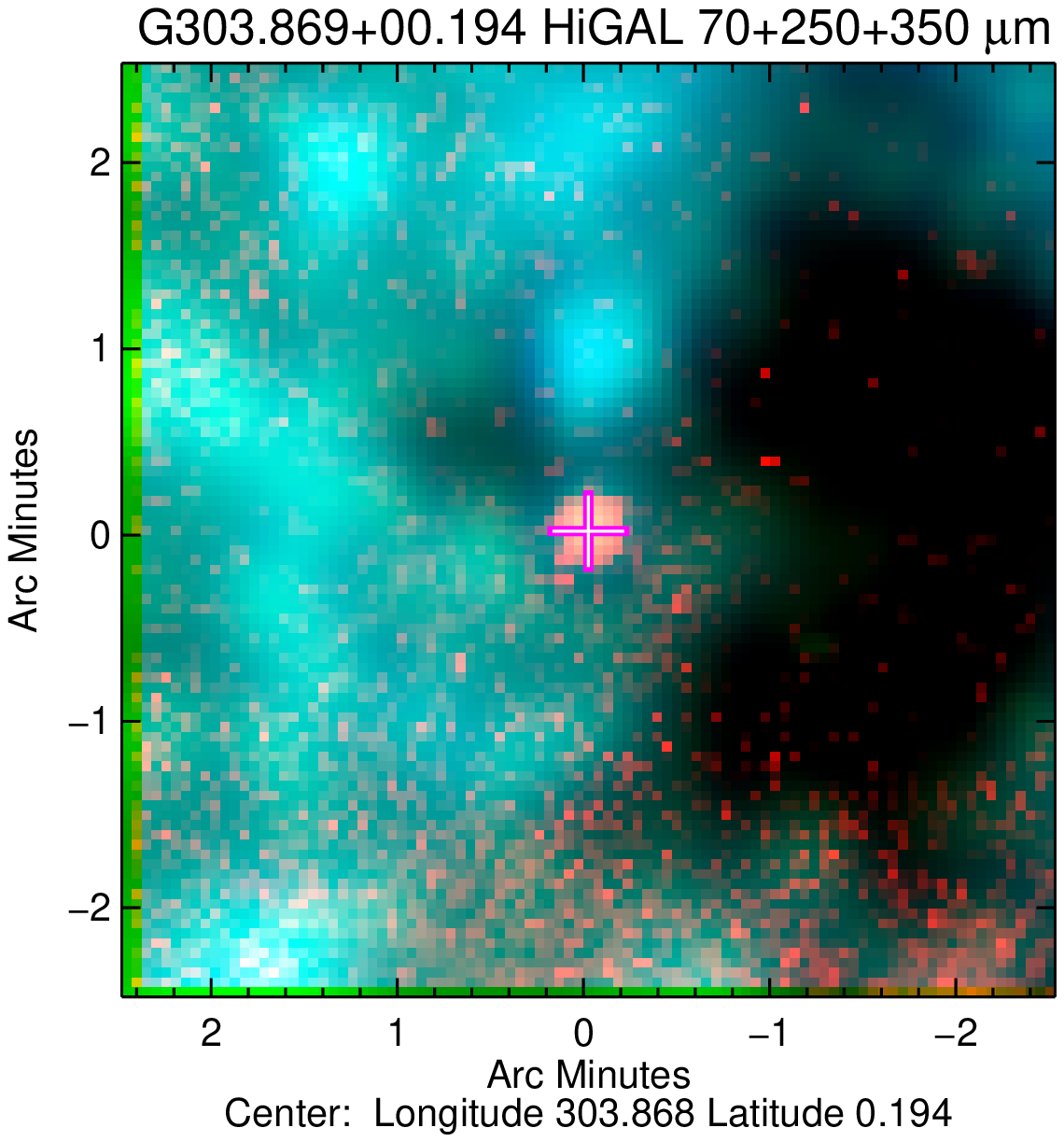}\\
\includegraphics[width=0.33\textwidth, trim= 0 0 0 0]{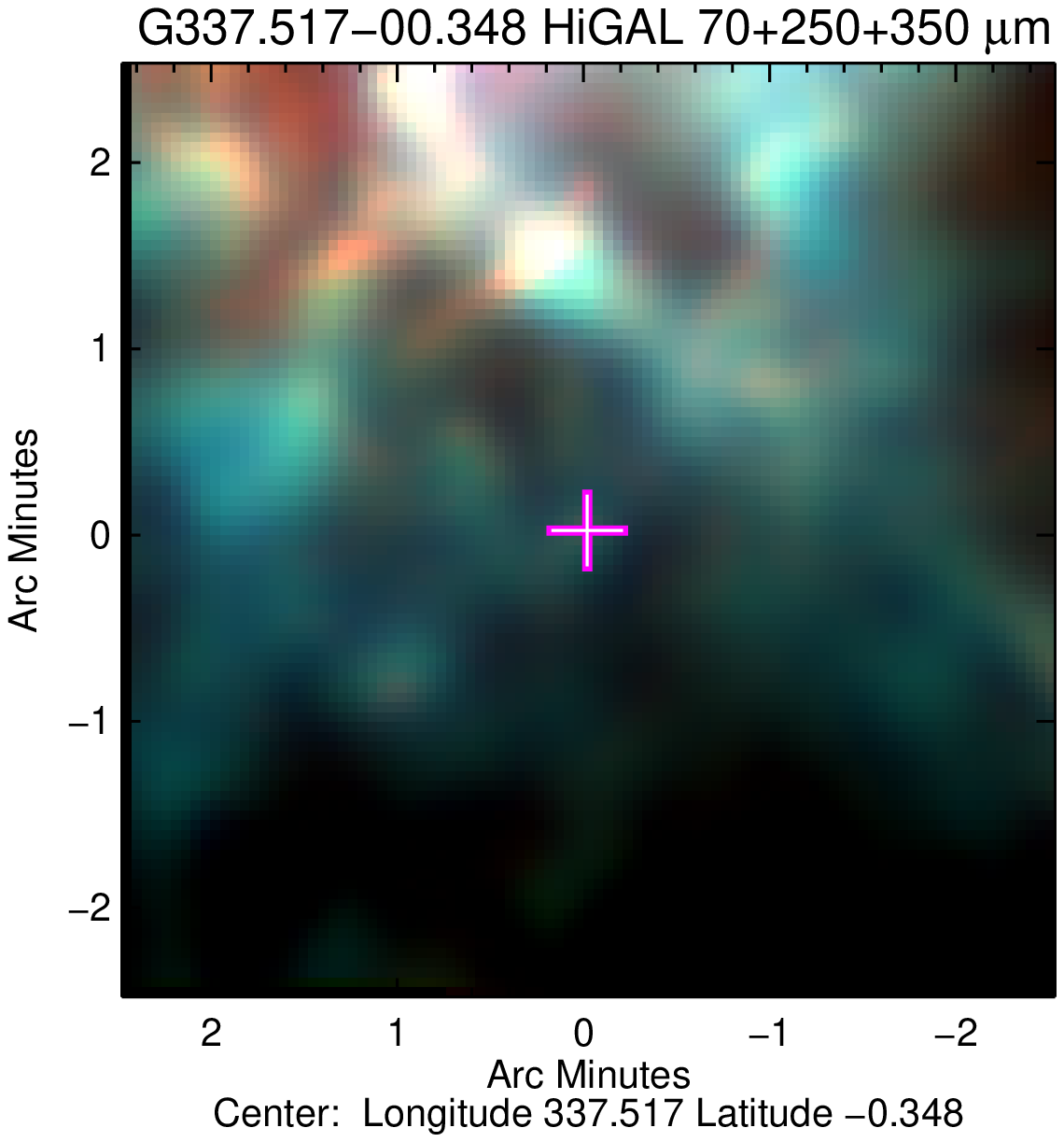}
\includegraphics[width=0.33\textwidth, trim= 0 0 0 0]{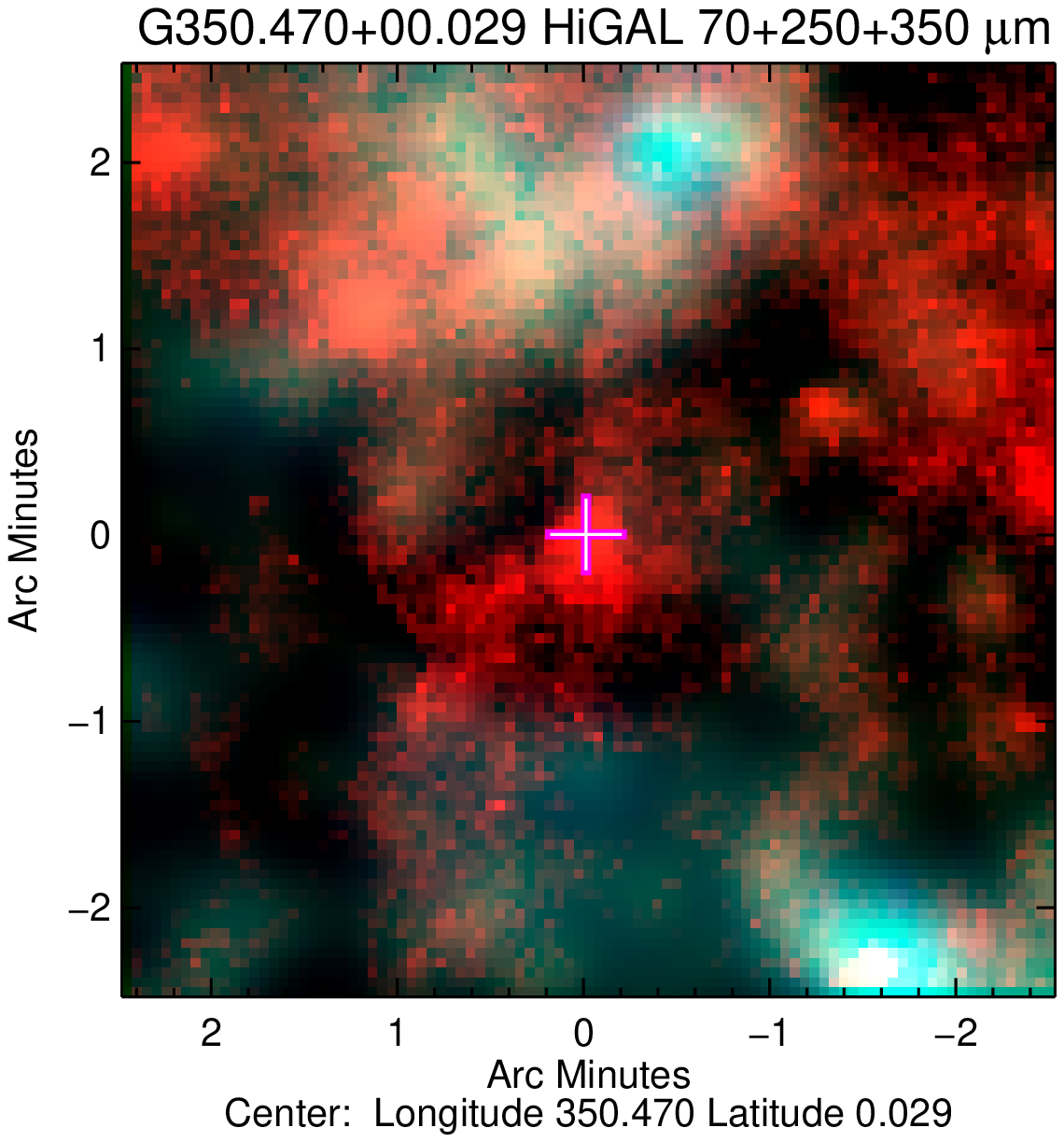}
\includegraphics[width=0.33\textwidth, trim= 0 0 0 0]{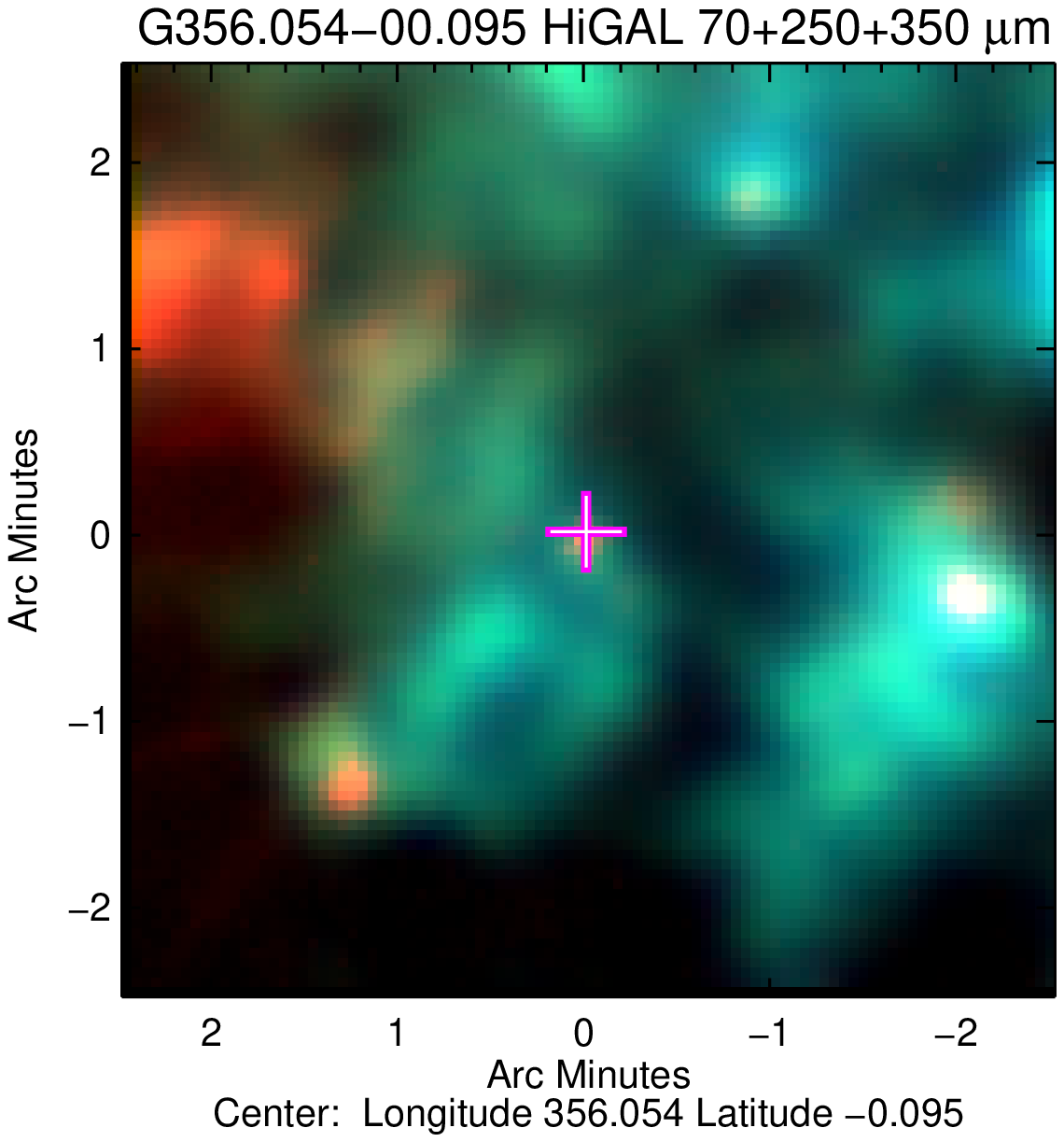}

\caption{\label{fig:higal_images} Three-colour composite images produced from the publicly available Hi-GAL data. These images use the 70, 250 and 350-\mum\ filters, which are coloured red, green and blue, respectively. The images are $5\arcmin \times5\arcmin$ in size and are centred on the methanol-maser position, which is indicated by the magenta cross.}

\end{center}
\end{figure*}

There are seven methanol masers that have not been associated with 870-\mum\ emission (i.e., SNR $> 3$). The catalogue names and derived parameters for these sources are given in Table\,\ref{tbl:ATLASGAL_dark_sources}. To investigate the nature of these methanol masers in more detail we have produced composite images from the publicly available level-2.5 tiles of the Hi-GAL open-time Key Project (\citealt{molinari2010}) obtained from the \textit{Herschel} Science Archive.\footnote{http://herschel.esac.esa.int/Science\_Archive.shtml. The high-pass-filtered maps have been used that have been calibrated and reduced with SPG v10.3.0 and combine the parallel and orthogonal scans.} In Fig.\,\ref{fig:higal_images} we present  three-colour images using the 70, 250 and 350\,\mum\ filters for six of these masers (MMB293.723$-$01.742 lies outside the HiGAL latitude range).

All but one of these masers are associated with compact 70-\mum\ emission. Only MMB337.517$-$003.348 appears to be dark at 70\,\mum\ but this source is at a distance of 17.1\,kpc and so this may be a sensitivity issue and, indeed, the estimated upper limit derived for its clump mass is poorly constrained (\mclump\ $< 154$\,\msun)). However, none is associated with any compact emission at longer wavelengths, which suggests that they are significantly warmer and less embedded than the rest of the methanol-maser population. There is some evidence of diffuse 350-\mum\ emission coincident with MMB014.521+00.155 and MMB356.054$-$00.095.  These are the same two sources towards which we find weak, diffuse 870-\mum\ emission (see Sect.\,3.1); however, the absence of any compact emission coincident with the position of the masers suggests that this is unlikely to be directly associated with these sources. 
The lack of any obvious 350-\mum\ emission probably means that deeper 870-\mum\ observations will not help and that these sources are likely to be associated with significantly lower dust column densities than the rest of the sample. 

We considered the possibility that these methanol masers may be spurious detections or the result of sidelobe contamination.  However, given that 5 of them are coincident with compact 70-\mum\ sources makes this unlikely. Another possibility is that the lack of contrast seen in the longer wavelength images could be a result of significant contamination from foreground material along the line of sight towards these sources. The combination of weak 350- and 870-\mum\  emission and compact 70-\mum\ emission is a little unexpected and suggests that these methanol masers are somewhat different from the rest of the population, however, further speculation on their nature requires additional data.  

\subsection{Galactic distribution}

\begin{figure*}
\begin{center}

\includegraphics[width=0.49\textwidth, trim= 20 10 20 20]{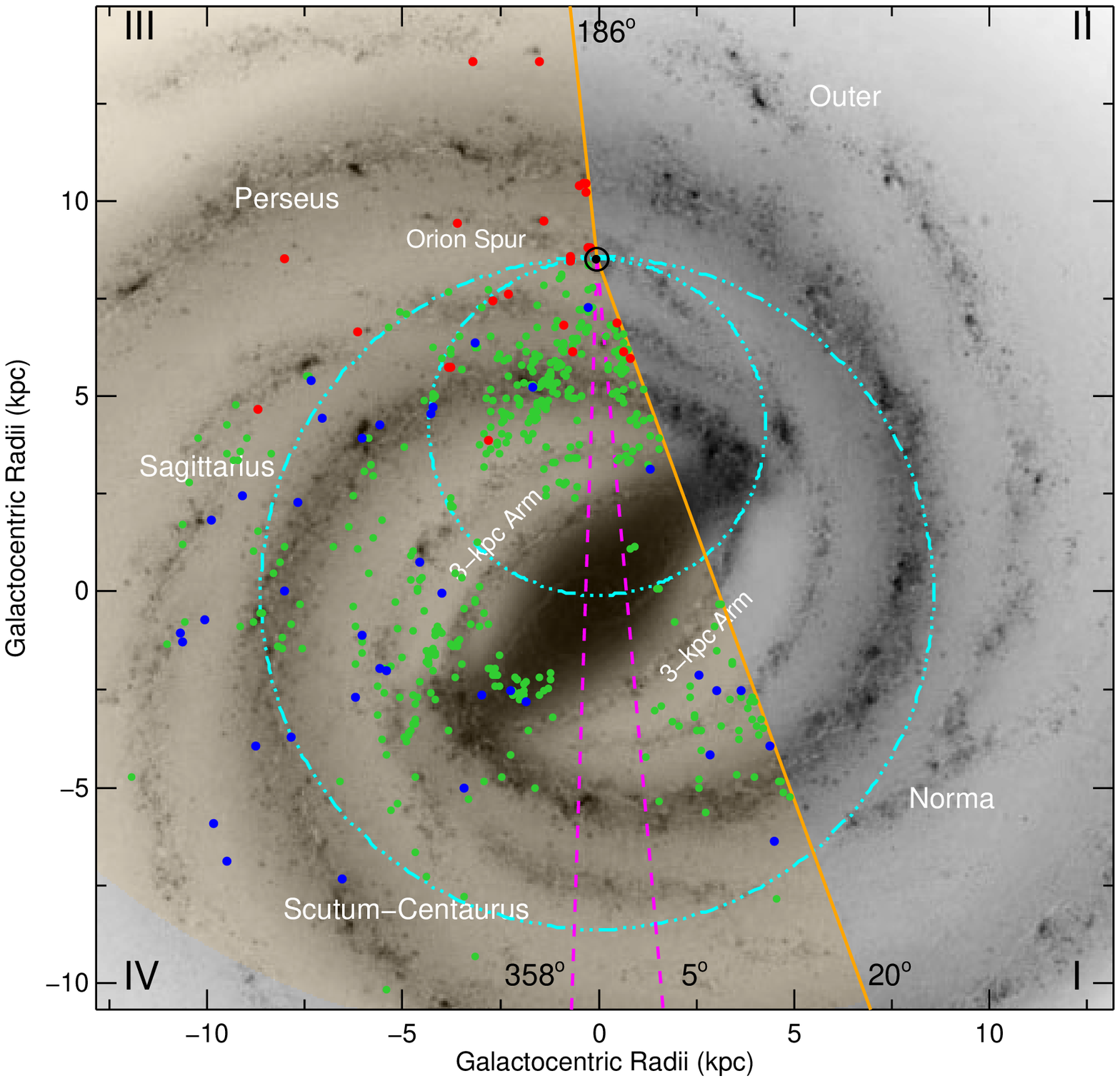}
\includegraphics[width=0.49\textwidth, trim= 20 10 20 20]{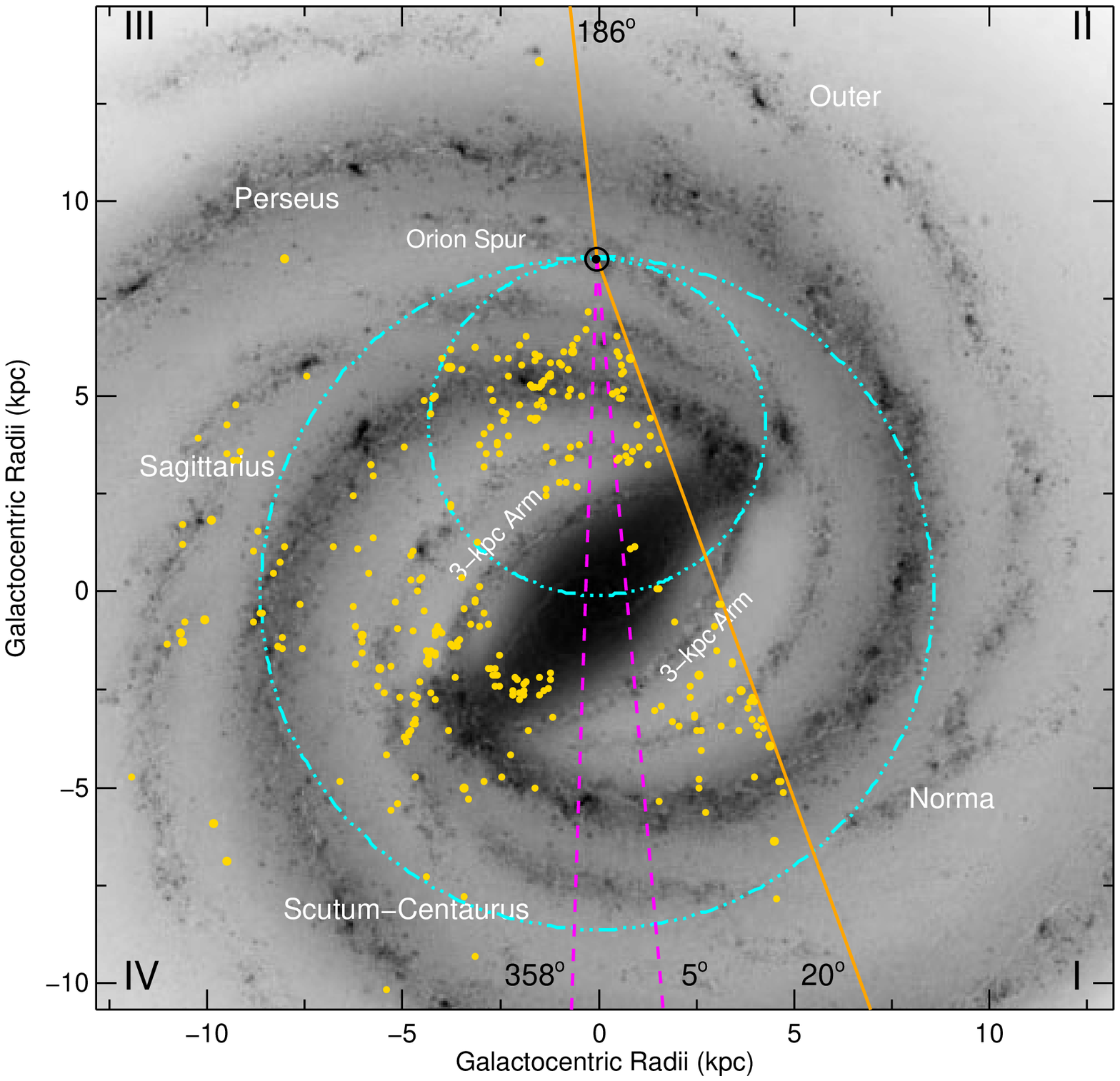}

\caption{\label{fig:galactic_distribution} Left panel: Galactic distribution of all masers associated with \submm\ clumps. The ATLASGAL-MMB matches are shown by green filled circles while the blue and red filled circles show the results of the deep and shallow observations discussed in this paper. The orange shaded area indicates the region of the Galactic plane covered by the MMB survey to a distance of 20\,kpc, within which the survey is complete for methanol masers with luminosities $>$1000\,Jy\,kpc$^2$. Right panel:  methanol maser associations with clump masses $>$ 1000\,\msun\ and maser luminosities $>$1000\,Jy\,kpc$^2$ for which our sample is complete to a distance of 20\,kpc. The background image is a schematic of the Galactic disc as viewed from the Northern Galactic Pole (courtesy of NASA/JPL-Caltech/R. Hurt (SSC/Caltech)). The Sun is located at the apex of the wedges and is indicated by the $\odot$ symbol. The smaller of the two cyan dot-dashed circles represent the locus of tangent points, while the larger circle traces the solar circle. The spiral arms are labeled in black and Galactic quadrants are given by the roman numerals in the corners of the image. The magenta line shows the innermost region towards the Galactic centre where distances are not available. } 

\end{center}
\end{figure*}

In Fig.\,\ref{fig:galactic_distribution} we present a schematic diagram of the Milky Way that includes many of the key elements of Galactic structure, such as the location of the spiral arms and the Galactic long and short bars. In the left panel we show the positions of all methanol masers associated with \submm\ continuum sources, for which a distance is available, while in the right panel we include only those matches above our completeness criteria (i.e., \mclump\ $>$ 1000\,\msun\ and $L_{\rm{MMB}}$ $>$1000\,Jy\,kpc$^2$). 

As discussed in Paper\,I there is a good correlation of the positions of the methanol masers with the \Sag\ and \Scu\ spiral arms, with the intersection of the end of the southern part of the long bar and the start of the \Per\ arm. However, the correlation between the parts of the \Per\ and Outer arms located in the third quadrant, which are relatively nearby compared to the other features mentioned, is significantly poorer. 
We considered that this might be the result of the limited latitude range of the MMB survey (i.e., $|b| < 2\degr$) and the deviation of these arms from the Galactic mid-plane to more negative values due to the Galactic warp. However, the Red MSX Source survey (\citealt{lumsden2013}), which covers a larger range in Galactic latitudes (i.e., $|b| < 5\degr$), also found a low number of MYSOs and compact \hii\ regions associated with the arms located in the third quadrant (\citealt{urquhart2014a}). This suggests that little massive star formation is taking place in the outer parts of the Galaxy compared to the inner Galaxy or that star formation in spiral arms is patchy and/or intermittent (\citealt{urquhart2014a}).

\begin{figure}
\begin{center}
\includegraphics[width=0.49\textwidth, trim= 0 0 0 0]{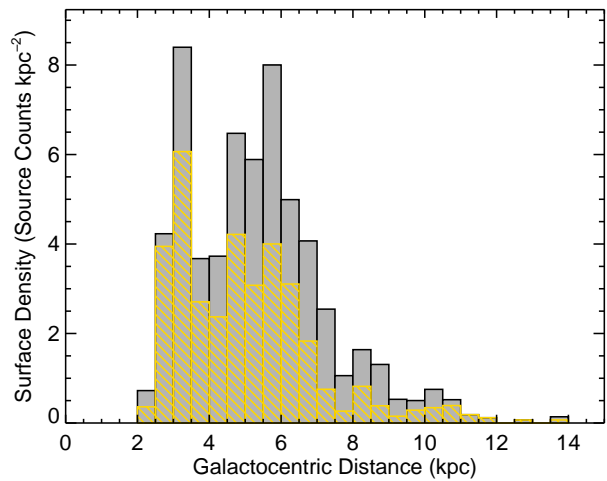}

\caption{\label{fig:rgc_distribution} Surface density distribution of all MMB-associated clumps located in the southern Galactic plane (i.e., $186\degr < \ell < 358\degr$) and for which a distance is available (grey histogram) and the subsample of these that are above the two completeness criteria (yellow hatched histogram); these two samples contain 429 and 242 sources, respectively. The bin size is 0.5\,kpc. }

\end{center}
\end{figure}

In Fig.\,\ref{fig:rgc_distribution} we present the surface density of the whole sample of methanol-maser associated clumps and the subsample above the completeness criteria, as a function of Galactocentric distance. 
The peaks at 3, 5 and 6\,kpc correspond to the far 3-kpc arm, the near section of the \Scu\ arms, and the end of the Galactic long bar, respectively. Although there is a good correlation between the methanol-maser associated clumps and the \Sag\ arm, the latter is not a prominent feature in the Galactocentric distribution, where we would expect to see a peak at $\sim$10\,kpc.  Although there are a number of massive star-forming regions associated with the \Sag\ arm, the star formation per unit area is relatively modest when compared to the segments of the spiral arms located inside the Solar circle.  This is in stark contrast to the high surface density of MYSOs and compact \hii\ regions reported by \citet{urquhart2014a} found to be associated with the \Sag\ arm and its association with RCW42, NGC3603 and G282.0$-$1.2 (Carina), which are some of the most active star-forming complexes in the Galaxy.

\begin{figure*}
\begin{center}
\includegraphics[width=0.49\textwidth, trim= 0 0 0 0]{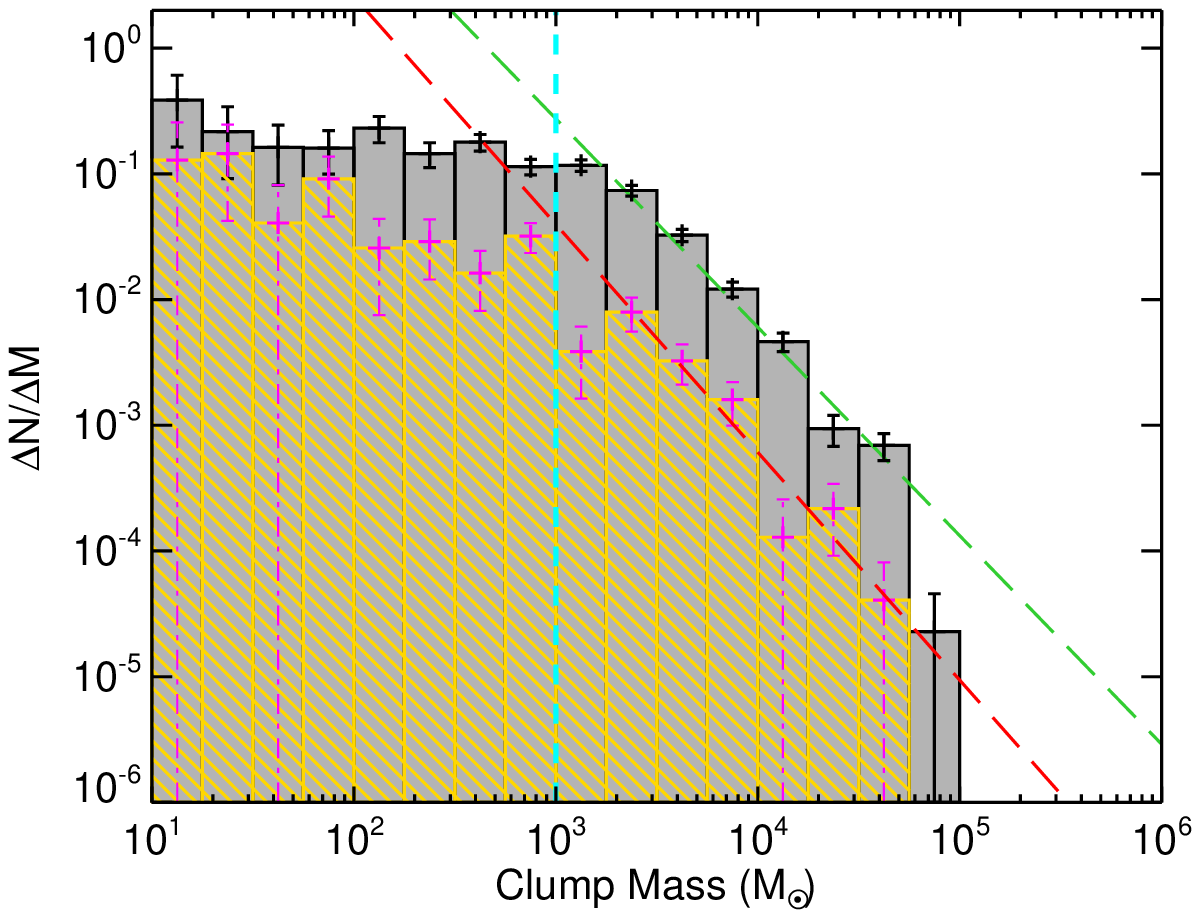}
\includegraphics[width=0.49\textwidth, trim= 0 0 0 0]{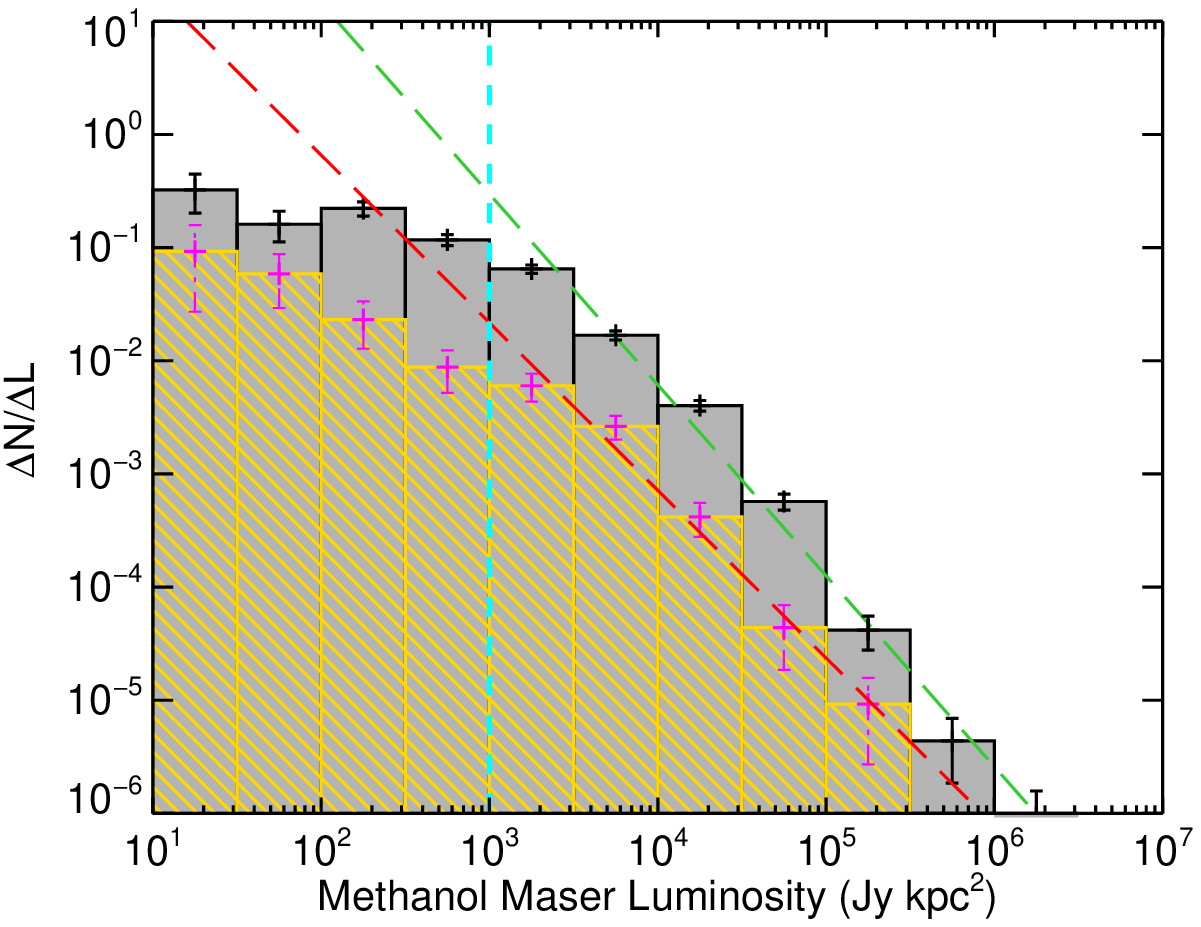}

\caption{\label{fig:mass_radius_dndm_histogram} Differential clump-mass and maser-luminosity distributions for all maser-associated clumps located in the inner (grey) and outer Galaxy (yellow). The inner- and outer-Galaxy mass functions have fitted power-law slopes of $-$1.66$\pm$0.15 and  $-$1.81$\pm$0.39, respectively. The inner- and outer-Galaxy luminosity functions have fitted slopes of $-$1.69$\pm$0.11 and  $-$1.49$\pm$0.12. The uncertainties are estimated using \poi\ statistics and the bin size used is 0.25\,dex and 0.5\,dex for the clump-mass and maser-luminosity distributions, respectively.} 

\end{center}
\end{figure*}

We find that approximately 10\,per\,cent of all methanol masers are located in the outer Galaxy, while the corresponding fraction of MYSOs and \hii\ regions is approximately twice this value. If we assume that the MYSOs and \hii\ regions are a more reliable indicator of the presence of high-mass star formation, then the frequency with which methanol masers are associated with massive star-forming regions is, for some reason, lower in the outer Galaxy.  While it is unclear what the underlying cause of this might be, the physical and environmental conditions in the outer Galaxy are very different, with weaker stellar gravity, lower H{\sc i} density, less intense UV radiation fields, smaller cosmic-ray flux (e.g., \citealt{bloemen1984}) and lower metallicity (e.g., \citealt{rudolph1997}) than the inner Galaxy. It is also likely that the inner- and outer-Galaxy parts of the spiral arms are different from each other (e.g., \citealt{benjamin2005}) with different star-formation mechanisms. The entry shock experienced by the interstellar medium (ISM) gas entering a spiral arm should only exist inside the co-rotation radius, where there is a differential velocity between the spiral pattern speed and the orbital rotation speed of the Galactic ISM. Outside this radius (thought to be just beyond the solar circle in the Milky Way; \citealt{lepine2011}), supernovae may be the dominant mechanism determining the state of the ISM and, hence, star formation (e.g., \citealt{kobayashi2008}). 

A low density of 6.7-GHz methanol masers located between $\ell=$ 290\degr\ and 305\degr\ has been previously noted by \citet{breen2011_methanol}. These authors also reported a significant decrease in the detection rate of 12.2-GHz methanol masers, which is half that found for the rest of study (i.e., 290\degr\ $< \ell <$ 10\degr). A similar low density of both 6.7- and 12.2-GHz methanol masers has been found for the Large Magellanic Cloud (LMC), which has approximately 5 times fewer methanol masers, relative to its star-formation rate, than the Milky Way, . This has been attributed to the lower metallicity found in the LMC (\citealt{green2008,ellingsen2010}) and \citet{breen2011_methanol} speculate that this may also explain the lower density of methanol masers in the outer Galaxy part of their survey. We can test this theory by comparing the methanol-maser luminosity and clump-mass functions for the inner and outer Galaxy populations. 

\subsubsection{Comparing the inner and outer Galaxy samples}

Although the molecular clouds located in the outer Galaxy may only contribute 20 per\,cent of the Galactic massive star formation, they provide a useful laboratory in which to study star formation under very different initial conditions. Furthermore, there are observational advantages to observing sources in the outer Galaxy, e.g., kinematic distances are unambiguous and there is reduced diffuse background emission and so source confusion is minimised. For these reasons a number of studies of the outer Galaxy population of molecular clouds have been conducted (e.g., {\citealt{wouterloot1989,may1997,nakagawa2005}). These surveys have identified a significant number of molecular clouds and find that the clouds are typically smaller, less massive and have narrower line widths than those located within the inner Galaxy (e.g., \citealt{dame1986,solomon1987}). However, the mass spectrum for inner and outer Galaxy clouds is found to be similar, which suggests that the cloud formation mechanism is independent of the Galactocentric radius, or that the mass spectrum is independent of the mechanism. 

The left panel of Fig.\,\ref{fig:mass_radius_dndm_histogram} shows the clump-mass functions for the inner- and outer-Galaxy samples of methanol-maser associated clumps.  Although the outer-Galaxy clumps are typically less massive than those located in the inner Galaxy, due to the smaller sample size, these two populations have fitted power-law slopes that are in excellent agreement, with values of $\sim$$-$1.8, similar to that found in a number of other studies of both molecular clouds and dense clumps.\footnote{Typical values for GMCs are between  $-$1.5 and $-$1.8  (e.g., \citealt{solomon1987, heyer2001, roman-duval2010}) and between $-2.0$\ to $-2.3$ for dense cores (e.g., \citealt{williams2004, reid2005,beltran2006,tackenberg2012}).} We note that the uncertainty on the slope of the outer Galaxy sample is large.  The good agreement in the slopes indicates that the mechanisms regulating the mass distributions of both clouds and the dense clumps within them are relatively invariant to Galactic location.  

The right panel of Fig.\,\ref{fig:mass_radius_dndm_histogram} shows the methanol-maser luminosity functions for the inner and outer Galaxy. Comparing these two distributions, we find that they again have similar slopes ($\sim$$-$1.6). That the slope is a little shallower than found for the clump mass function is to be expected, since the luminosity increases much more quickly than the mass for the embedded source (see Fig.\,8). The similarity between the inner- and outer-Galaxy luminosity functions suggests that basic conditions required for the maser emission, such as optical path length, pumping mechanism and volume densities, are also likely to be invariant to Galactic location. 

\begin{figure}
\begin{center}
\includegraphics[width=0.49\textwidth, trim= 0 0 0 0]{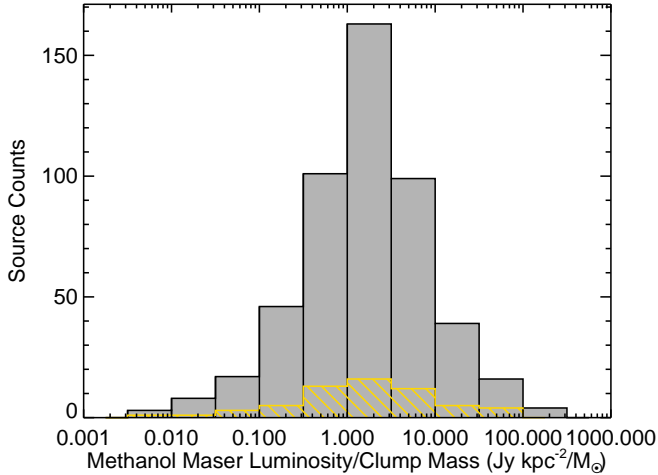}

\caption{\label{fig:inner_outer_ratio} $L_{\rm{mmb}}$/\mclump\ ratio for the inner- and outer-Galaxy samples, shown in grey and yellow hatching, respectively. The inner- and outer-Galaxy samples consist of 466 and 59 sources, respectively, and the bin size is 0.5\,dex. }

\end{center}
\end{figure} 

The similarities of the clump mass and maser luminosity functions means that there is also no significant difference in the maser-luminosity to clump-mass ($L_{\rm{mmb}}$/\mclump) ratio (see Fig.\,\ref{fig:inner_outer_ratio}). The mean values for the inner- and outer-Galaxy samples are 6.2$\pm$0.9 and 6.3$\pm$1.6, respectively, and a \KS\ (KS) test is unable to reject the null hypothesis that these two samples are drawn from the same parent population ($r=0.92$). At first glance this is somewhat surprising, given that the metallicity is approximately a factor of three lower in the outer Galaxy (\citealt{lepine2011}). However, it is also likely that the lower metallicity will result in a corresponding lower dust column density and so the clump masses are probably underestimated. If the underlying $L_{\rm{mmb}}$/\mclump\ ratio were lower in the outer Galaxy due to this, it would be consistent with the lower metallicity hypothesis speculated upon by \citet{breen2011_methanol} to explain the lower fraction of methanol masers found there.

\subsection{Evidence for an evolutionary sequence}

There have been a number of attempts to use the presence or absence of the various kinds of interstellar masers (e.g., class\,I and II methanol masers, OH and H$_2$O),  their properties and those of their host clump to define an evolutionary sequence for massive stars (e.g., \citealt{ellingsen2007,ellingsen2011}). This evolutionary framework is based on the stronger positional correlation found between more luminous 6.7- and 12.2-GHz methanol masers and water masers, with OH masers and radio-continuum emission (\citealt{breen2010,breen2011_methanol}), which are typically associated with the most evolved embedded stages of the massive-star formation process. In this subsection we will use the derived clump properties and methanol-maser luminosities to further explore this evolutionary model.

As a consequence of the large range of possible final stellar masses (and larger range of bolometric luminosities)  for OB stars, the luminosity alone is not sufficient to determine the evolutionary stage of the embedded object. However, we should expect the $L_{\rm{mmb}}$/\mclump\ ratio to increase as the embedded protostellar object evolves through the various evolutionary stages (i.e., hot molecular core $\rightarrow$\ MYSO $\rightarrow$\ UC \hii\ region phases). This is analogous to the  mass-luminosity model used by \citet{saraceno1996} to study the evolution of low-mass stars and developed by \citet{molinari2008} to track the evolution of high-mass stars embedded in massive clumps. In Paper\,I we found a correlation between the luminosities of 6.7-GHz methanol masers and the bolometric luminosity of the embedded objects and therefore the $L_{\rm{mmb}}$/\mclump\ and $L_{\rm{bol}}$/\mclump\ ratios should be equivalent.  Also physically, if the masers are saturated, we expect the maser output to be linearly dependent on the photon pump rate. The basic premise of this model is that the bolometric luminosity increases dramatically during the embedded star's early evolution while the mass of the clump is relatively unaffected. $L_{\rm{bol}/\rm{mmb}}$/\mclump\ is therefore expected to increase as the embedded source evolves towards the main sequence. This model has been reasonably successful in tracing the evolution of massive protostars (e.g., \citealt{molinari2008, giannetti2013, urquhart2014b}). 

If the evolutionary model constructed from the maser studies is correct, we should find that the $L_{\rm{mmb}}$/\mclump\ ratio increases as the embedded object evolves towards the main sequence. We can test this by comparing the $L_{\rm{mmb}}$/\mclump\ ratio for a sample of methanol masers known to be associated with some of the more evolved embedded stages. For this we use the sample of methanol masers found to be associated with MYSOs and UC\,\hii\ regions identified by the RMS survey (\citealt{lumsden2013,urquhart2014b}). To make sure that these associations are genuine, we require that the angular offset between the methanol maser and matched MYSO or \hii\ region is less than 2\arcsec\ and place a cut on the heliocentric distance of 10\,kpc; these criteria ensure that the physical separation between the methanol maser and MYSO/\hii\ region is $<0.1$\,pc, the expected size of a protostellar core. Fig.\,\ref{fig:physical_offset} shows the physical separation between the methanol masers and their associated MYSO or UC\,\hii\ region counterparts. The distribution is strongly peaked towards physical separations of a few hundredths of a parsec and we can therefore be reasonably confident they are tracing the same embedded object.

\begin{figure}
\begin{center}
\includegraphics[width=0.49\textwidth, trim= 0 0 0 0]{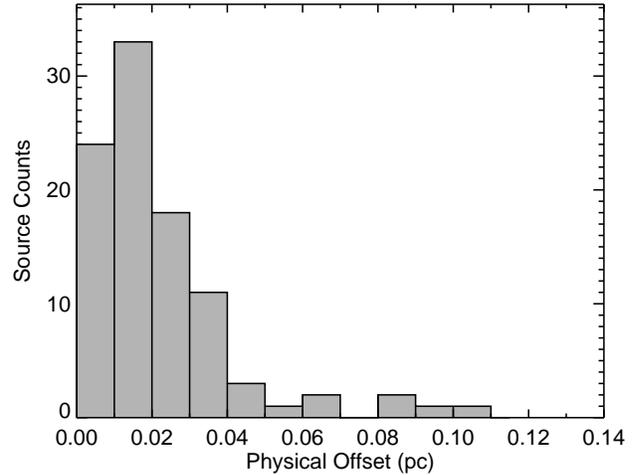}

\caption{\label{fig:physical_offset} Physical offset between methanol masers and their associated MYSO and UC\,\hii\ regions (i.e., angular projected separation $<$2\arcsec). In total, 113 associations have been identified (\citealt{urquhart2014b}) and these have an average separation of 0.023$\pm$0.002\,pc and median value of 0.017\,pc. The bin size used is 0.01\,pc.} 

\end{center}
\end{figure}

Fig.\,\ref{fig:mass_lum_ratio_histogram} shows the distribution of the $L_{\rm{mmb}}$/\mclump\ ratio for the whole sample and for the subsample of methanol masers associated with MYSOs and UC \hii\ regions. If the methanol-maser luminosity were a good tracer of the evolutionary stage we would expect to see a clear difference in the distribution of these subsamples, with higher values for the more evolved stages. However, this is manifestly not the case and the shapes of the two distributions are almost indistinguishable. The peaks of the distributions agree within their standard errors (6.9$\pm$1.2 and 10.2$\pm$2.6 for the methanol maser only clumps and MYSO/\hii\ region associated subsamples, respectively) and a KS test is unable to reject the null hypothesis that these two samples are drawn from the same parental distribution ($p=0.44$). 

\begin{figure}
\begin{center}
\includegraphics[width=0.49\textwidth, trim= 0 0 0 0]{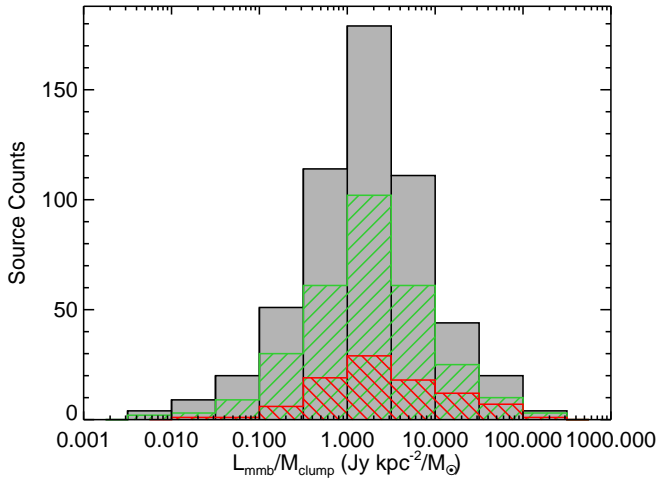}

\caption{\label{fig:mass_lum_ratio_histogram} The $L_{\rm{mmb}}$/\mclump\ ratio for the whole sample of methanol-maser associated clumps (grey), the MYSO- and UC \hii -region associated subsample (red) and unassociated methanol-maser clumps (green). In this plot, we combine the MYSO- and \hii -region associated samples, as both stages have been found to be close to the end of the main accretion phase (\citealt{urquhart2014b}). The bin size used is 0.5\,dex.} 

\end{center}
\end{figure}

From this, we conclude that the $L_{\rm{mmb}}$/\mclump\ ratio is insensitive to the current evolutionary stage of the embedded object. This is a little surprising given the success of the $L_{\rm{bol}}$/\mclump\ ratio and that clear differences in this ratio have been found between methanol-maser-only clumps and MYSO/\hii -region associated clumps (\citealt{urquhart2014b}). On average, the $L_{\rm{bol}}$/\mclump\ ratio for the MYSO- and UC\,\hii -region associated clumps is twice that found for the maser-only clumps. This suggests that, while the bolometric luminosity increases as the source evolves, the maser luminosity remains relatively constant, assuming that the clump mass does not change significantly during the course of the massive star's evolution. 

\begin{figure}
\begin{center}
\includegraphics[width=0.49\textwidth, trim= 0 0 0 0]{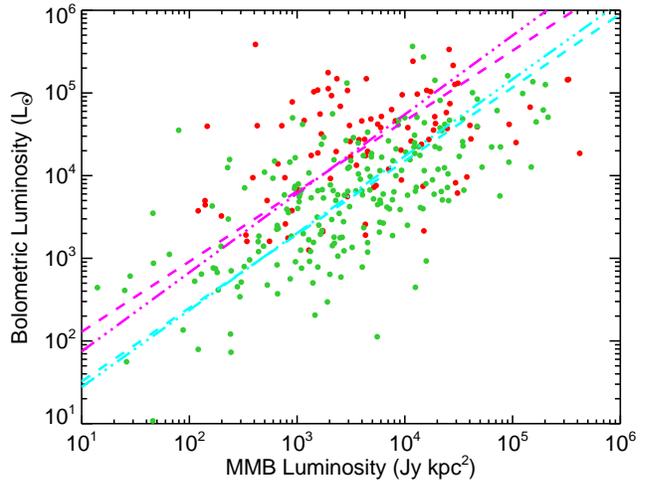}

\caption{\label{fig:luminosity_scatter} Methanol-maser luminosity versus bolometric luminosity. The red and green circles show the distribution of the MYSO and UC\,\hii -region associated methanol masers, while green circles give the luminosities of methanol-maser-only clumps. The partial-Spearman correlation coefficients for the MYSO/\hii -region and maser-only clumps are 0.32 and 0.47, respectively. The magenta and cyan lines show the results of a power-law fit to the maser-MYSO/\hii -region associated clumps (slope=0.96$\pm$0.24) and the maser-only clumps (slope=0.93$\pm$0.08), respectively. The long-dashed lines are fits to all sources in each sample while the dot-dashed lines are fits to a distance limited subset (sources between 3 and 5\,kpc --- the slopes are determined from the distance-limited sample, although there is no significant difference between them).} 

\end{center}
\end{figure}

Fig.\,\ref{fig:luminosity_scatter} shows the methanol-maser and bolometric luminosities of the MYSO/\hii\ regions and of the methanol-maser-only associated clumps (\citealt{urquhart2014b}). The distribution of the two samples in this plot supports the conclusion that the maser luminosity is does not change significantly as the embedded source evolves and casts further doubt on the maser luminosity being a useful diagnostic of evolution, contrary to many of the maser-based evolutionary models (e.g. \citealt{breen2010,breen2011_methanol,ellingsen2011,ellingsen2013}). One explanation for the disgreement is that the latter studies have interpreted differences in maser luminosity as indicative of age, while we have combined the luminosity and clump mass to take account of the embedded object's dependence on both age and mass. Another possible explanation is that the we may have underestimated the bolometric luminosities, which are estimated by scaling the 70\,\mum\ Hi-Gal luminosity (see Sect. 4.6.2 of \citealt{urquhart2014b} for details), however, this is considered less likely. A more detailed analysis is beyond the scope of the current work but should be undertaken in the future to resolve these inconsistencies.

\section{Summary and conclusions}

We present the results of a set of 870-\mum\  continuum observations made towards 77 class\,II 6.7-GHz methanol masers identified by the Methanol Maser multiBeam (MMB) survey (\citealt{caswell2010b}). Using the LABOCA instrument on the APEX telescope to map the thermal dust emission round each methanol maser ($\sim$8\arcmin\ diameter). Analysis of these maps has resulted in the identification of $\sim$300 compact sources and has matched 70 of the methanol masers targeted with \submm\ emission. Combining with the analysis of 630 methanol masers associated with dust emission from a previous study provides a complete census of \submm\ observations towards the entire MMB catalogue, which consists of 707 methanol masers. We estimate the physical properties of the methanol masers and their host clumps and examine their nature and Galactic distribution. Our main findings are as follows:

\begin{enumerate}

\item When combined with the methanol masers previously studied, we find that the association rate is 99\,per\,cent of the full MMB catalogue. Evaluation of the derived dust and maser properties leads us to conclude that the combined sample represents a single population and likely to be tracing the same phenomenon. Furthermore, we find typical clump masses of a few 10$^3$\,\msun\ and that all but a handful satisfy the mass-size criteria required for massive star formation. This study provides the strongest evidence of the ubiquitous association of methanol masers with compact dense clumps and shows them to be excellent signposts for identifying massive star forming regions. 

\item We examined the publicly available Hi-GAL 70, 160, 250, 350 and 500\,\mum\ images towards the six of the seven methanol masers not currently associated with dust emission. We find weak emission coincident with two sources but no evidence that any of the others are associated with any significant cold dust emission. The absence of compact \submm\ continuum sources towards these methanol masers is likely to be real and not due to a lack of sensitivity of the observations. These methanol masers do appear to be significantly different from the rest of the population.

\item Comparing the ratio of inner and outer Galaxy methanol masers to the ratio of MYSOs and \hii\ regions, we find a significantly lower fraction of methanol masers located in the outer Galaxy. Approximately 20\,per\,cent of the MYSO and \hii\ regions are found in the outer Galaxy compared to only 10\,per\,cent of the methanol masers. The lower metallicity environment of the outer Galaxy may offer an explanation, however, further studies are required to confirm this.

\item We compare the clump-mass and maser-luminosity functions for the inner- and outer-Galaxy populations and find that although the outer-Galaxy associations are less massive and have lower maser luminosities than typically found in the inner Galaxy, the slopes of the two distributions are very similar. It seems likely that the clump-forming efficiency is invariant to Galactic location in much the same way that the mechanism for cloud formation appears to be.

\item We use the ratio of the methanol-maser luminosity to clump mass to test the hypothesis that the maser luminosity is positively correlated with the evolutionary stage of the embedded source. We compare this ratio for a sample of methanol-masers associated with MYSOs and UC\,\hii\ regions with that for more isolated methanol masers and find no evidence for a link with evolution. From this we conclude that the maser luminosity is dominated by the bolometric luminosity of the embedded source and is not a good indicator of its evolutionary stage.

\end{enumerate}

\section*{Acknowledgments}

We would like to thank the staff of the APEX telescope for their support during the observation, particularly Claudio Agurto  and Felipe Mac Auliffe for their assistance with the data reduction. We would also like to thank Simon Ellingsen and Shari Breen for discussions on the maser evolution model and the referee Michael Burton for helpful comments and suggestions. This research has made use of the SIMBAD database operated at CDS, Strasbourg, France. This work was partially funded by the ERC Advanced Investigator Grant GLOSTAR (247078) and was partially carried out within the Collaborative Research Council 956, sub-project A6, funded by the Deutsche Forschungsgemeinschaft (DFG). 

\bibliography{rms}

\bibliographystyle{mn2e_new}

\end{document}